\documentclass[12pt]{article}
\usepackage[margin=1in]{geometry}
\usepackage{times}

\usepackage{fullpage, amsmath, amsthm, amsfonts, rotating}
\usepackage[round]{natbib}
\usepackage{booktabs}
\usepackage{bbm}
\usepackage{hyperref}
\usepackage{nameref}
\usepackage{verbatim}
\usepackage{natbib}
\usepackage{lmodern,xcolor}
\usepackage{graphicx}
\usepackage{amssymb}
\usepackage{epstopdf}
\usepackage{algorithm}
\usepackage{algorithmic}

\usepackage{bm}
\usepackage{comment}

\usepackage{caption}
\usepackage{subcaption}
\allowdisplaybreaks
\usepackage{tikz}

\newcommand{\bx}{\bm x}

\newcommand{\bbeta}{\bm \beta}

\newcommand{\btheta}{\bm \theta}

\newcommand{\logit}{\mbox{logit}}
\newcommand{\expit}{\mbox{expit}}

 \usepackage{booktabs}

\usepackage{tikzsymbols}
\usepackage{tikz}
\usetikzlibrary{shapes,decorations,arrows,calc,arrows.meta,fit,positioning,shadows}
\tikzset{
    every picture/.style=thick,
    -Latex,auto,node distance =1 cm and 1 cm,semithick,
    state/.style ={ellipse, draw, minimum width = 0.7 cm},
    point/.style = {circle, draw, inner sep=0.04cm,fill,node contents={}},
    bidirected/.style={Latex-Latex,dashed},
    el/.style = {inner sep=2pt, align=left, sloped}
}
\usepackage[rightbars, color]{changebar}
\cbcolor{blue}
\newcommand{\rev}[1]{%
  \cbcolor{white}
  \begin{changebar}
    #1
  \end{changebar}%
  }%

\usepackage{authblk}

\title{Toward a Principled Workflow for Prevalence Mapping Using Household Survey Data}
\author[1]{Qianyu Dong}
\author[2]{Yunhan Wu}
\author[1]{Zehang Richard Li} 
\author[2, 3]{Jon Wakefield}
\affil[1]{Department of Statistics, University of California, Santa Cruz}
\affil[2]{Department of Biostatistics, University of Washington}
\affil[3]{Department of Statistics, University of Washington}
\date{\today}
 
\begin{document}    
\maketitle
\abstract{
    Understanding the prevalence of key demographic and health indicators in small geographic areas and domains is of global interest, especially in low- and middle-income countries (LMICs), where vital registration data is sparse and household surveys are the primary source of information.
    Recent advances in computation and the increasing availability of spatially detailed datasets have led to much progress in sophisticated statistical modeling of prevalence. As a result, high-resolution prevalence maps for many indicators are routinely produced in the literature. 
    However, statistical and practical guidance for producing prevalence maps in LMICs has been largely lacking. In particular, advice in choosing and evaluating models and interpreting results is needed, especially when data is limited. Software and analysis tools are also usually inaccessible to researchers in low-resource settings to conduct their own analysis or reproduce findings in the literature.
    In this paper, we propose a general workflow for prevalence mapping using household survey data. We consider all stages of the analysis pipeline, with particular emphasis on model choice and interpretation. We illustrate the proposed workflow using a case study mapping the proportion of pregnant women who had at least four antenatal care visits in Kenya. Reproducible code is provided in the Supplementary Materials and can be readily extended to a broad collection of indicators.  
}

\section*{Statement of Significance}
We propose a principled workflow for estimating and mapping the prevalence of demographic and health indicators in small administrative areas using household survey data in low- and middle-income countries (LMICs). The workflow considers all stages of the data analysis pipeline and can be robustly adopted for a wide range of binary indicators. The aim of the workflow is to enable researchers and analysts in LMICs to carry out their own prevalence analysis according to their local needs and priorities. The workflow is illustrated through a case study mapping the proportion of pregnant women who had at least four antenatal care visits in Kenya, with reproducible code in the Supplementary Materials.

\section{Introduction}
Accurate estimation and tracking of key demographic and health indicators is critical for assessing health progress and disparities. 
National statistical institutions and global organizations have long recognized the need to produce disaggregated estimates at fine resolutions.  
For example, the United Nations General Assembly's 2030 Agenda for Sustainable Development \citep{sdgsWeb} established targets 
``to increase significantly the availability of high-quality, timely and reliable data disaggregated by income, gender, age, race, ethnicity, migratory status, disability, geographic location and other characteristics relevant in national contexts.'' 
The problem of estimating quantities of interest for small subpopulations based on limited data is known as small area estimation (SAE). SAE has a long history in the survey sampling literature. Traditional SAE approaches have been largely developed in the high-income country context and often utilize auxiliary administrative data to improve inference \citep{rao:molina:15}. Recent applications of SAE approaches in LMICs include work on mapping poverty \citep{molina2019small, chi2022microestimates}, child mortality \citep{mercer:etal:15, li:etal:19, wakefield:etal:19, wu:etal:21}, fertility \citep{saha2023small}, HIV prevalence \citep{wakefield2020small}, vaccination coverage \citep{dong:wakefield:21}. More recently, another line of research using geostatistical models has been adopted by major research organizations to map the prevalence of a number of important health indicators \citep[see e.g.,][]{diggle:giorgi:16, diggle:giorgi:19, osgood:etal:18, burstein:etal:18, utazi:etal:18, mayala:etal:19}. Geostatistical models theoretically enable prevalence estimation at a fine spatial resolution, e.g., $100$m grids, and they leverage the growing availability of spatially referenced covariates such as population density \citep{tatem2017worldpop} and night time light \citep{gaughan2019evaluating}. A recent review and comparison of the two types of approaches is provided in \citet{wakefield2020two}.

\rev{While rapid progress has been made in methodological developments of mapping prevalence in small areas, the existing literature often focuses on improving specific steps of the full SAE workflow. Guidance on conducting prevalence mapping from the beginning to the end has been largely lacking, especially for practitioners. 
}  
Two recent papers, \citet{tzavidis:etal:18} and \citet{giorgi2021model}, discuss the significance of having a coherent process across different stages of analysis when producing small area official statistics. 
However, scenarios where data is scarce remain challenging under both frameworks.
\citet{tzavidis:etal:18} examined the production of official statistics from the perspective of national statistical institutes. Their framework follows a traditional SAE approach and focuses on continuous outcomes, e.g., average income. \rev{The models for continuous outcomes are usually inappropriate for estimating the prevalence of categorical indicators, which is the focus of this paper.} In addition, their framework emphasizes using covariates from census or administrative microdata in order to construct synthetic estimates at fine geographic scale. In LMICs, the issues with data availability are quite different from those faced in high-income countries. Auxiliary datasets from census are usually unavailable, or of poor quality and are often not updated in a timely fashion. Thus, it is important to exploit the spatial dependence in the response variable that cannot be explained by the covariates that are available to the analyst. This aspect is largely ignored in \citet{tzavidis:etal:18}.  
On the other hand, \citet{giorgi2021model} provided a parallel set of guidance on developing prevalence models using geostatistical models, focusing on using high-quality spatially referenced covariates for point-level spatial prediction. 
In contrast with \citet{tzavidis:etal:18}, some key considerations in the traditional SAE methods, e.g., accounting for the sampling design of a survey, are mostly ignored. 
Notably, both \citet{tzavidis:etal:18} and \citet{giorgi2021model} were primarily written for statisticians capable of building and comparing sophisticated models. Translating the ideas to routine analytic procedures involving large numbers of indicators can be difficult for practitioners. 

In this paper, we aim to fill this gap by developing a principled and accessible workflow for common types of prevalence mapping tasks in data-limited contexts with fully worked-out examples.
Specifically, we focus on estimating the prevalence of binary indicators, i.e., the proportion of individuals affected by certain conditions.
We develop a `default' analysis pipeline with a series of models for prevalence mapping that can be adopted robustly for a wide range of indicators. 
Our proposed procedures acknowledge the sampling design explicitly and account for spatial dependence in the prevalence. The models are computationally scalable, making them feasible to implement in low-resource settings. Altogether, the workflow generates a robust set of prevalence estimates and insights that can form the basis for further extensions when bespoke analyses are needed. 

The rest of the paper is organized as follows. Section \ref{sec:data} discusses the common types of household survey data in LMICs. Section \ref{sec:model} describes the steps in the workflow.
Section \ref{sec:discuss} concludes with a discussion of the proposed workflow, active research areas, and open questions. 
We demonstrate the proposed workflow via a case study mapping the proportion of pregnant women who had at least four antenatal care visits using a Demographic and Health Survey (DHS) that was carried out in Kenya, in 2022. 

\section{Household survey data}\label{sec:data}

In many LMICs, household surveys are routinely conducted to collect information on health and demographic variables. Prominent examples include the DHS \citep{dhs}, the Malaria Indicators Survey (MIS) \citep{malaria2005malaria}, and the Multiple Indicator Cluster Surveys (MICS) \citep{khan2019multiple}. We consider the 2022 Kenya DHS \citep{knbs2023kenya} as our working example. The survey is based on a sampling frame derived from the 2019 Kenya Population and Housing Census, in which a total of $129,067$  enumeration areas (EAs), or clusters, were formed. In the 2022 Kenya DHS, the country is first stratified into rural and urban strata within each of the $47$ counties, which results in $92$ strata (Nairobi City and Mombasa are purely urban). Within each stratum, clusters are sampled independently using probability proportional to size (PPS) random sampling, with households being the size variable. This resulted in $1,692$ clusters being sampled across strata. Then $25$ households were selected from each cluster and members of the households were interviewed to provide information on health and demographic variables. The design weight for each individual is calculated as the reciprocal of the sampling probability for the individual. The final weight also includes a non-response adjustment.

\rev{Household surveys in LMICs, including DHS, MICS, and MIS, often follow similar sampling designs.} For ease of description, we refer to the principal administrative divisions used to define the stratification, e.g., the $47$ counties in Kenya, as the Admin-1 areas, and the next level of $300$ subdivisions as the Admin-2 areas. \rev{Note that the geographical stratification  is survey-specific and may be Admin-1 or Admin-2, depending on the country.} The number of administrative areas varies dramatically across countries. For example, there are $33$ Admin-2 areas in C\^ote d'Ivoire and $774$ in Nigeria. The sample size in most household surveys is designed to provide estimates of key indicators at the national and Admin-1 level only. Data are typically sparse at the Admin-2 level. 

The DHS program routinely calculates the national estimates of over $2,500$ indicators for each survey \citep{dhsguide}. The majority of these indicators are binary.
As a working example, we consider the prevalence of attending four or more antenatal care visits (ANC4+) among women who had a live birth or a stillbirth in the $5$ years prior to the survey. Antenatal care is a crucial component in reducing maternal and perinatal mortality. 
The World Health Organization (WHO) recommended at least four ANC visits with a qualified healthcare provider during pregnancy  \citep{world2002antenatal} and a global ANC4+ coverage target of $90\%$ by 2025, in order to meet the SDG goals on maternal mortality by 2030.
Several studies have examined the ANC4+ coverage in Kenya using past DHS surveys, see for example \citet{wairoto2020determinants} and \citet{macharia2022spatial}.
The national estimates of ANC4+ coverage in Kenya have been steadily increasing, from $44\%$ in 2009 to $66\%$ in 2022 \citep{knbs2023kenya}.

\section{The workflow} \label{sec:model}
\begin{figure}[!ht]
    \centering
    \includegraphics[width=\linewidth]{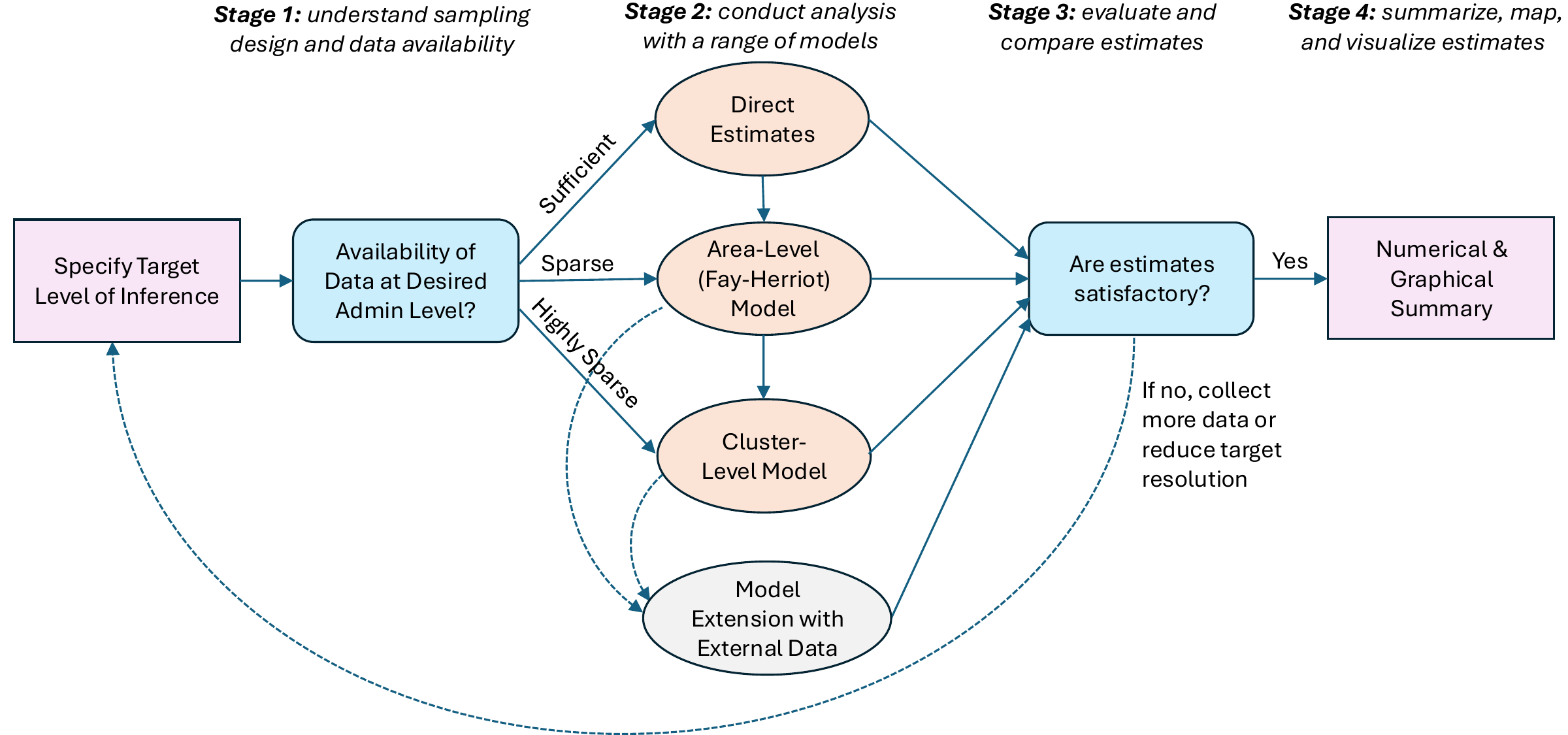}
    \caption{Workflow for prevalence mapping using household survey data.}
    \label{fig:workflow}
\end{figure}

Our prevalence mapping workflow starts by specifying a level of inference. Typically, household surveys powered to Admin-1 areas can robustly produce national and Admin-1 level estimates. Inference at finer resolutions relies more heavily on model assumptions, as we detail later. 
In this paper, we consider prevalence mapping at both Admin-1 and Admin-2. 
The same workflow follows for finer resolutions, though data sparsity will be more severe.
Our proposed workflow consists of four stages summarized in Figure \ref{fig:workflow}: 

\vspace{.2in}

    Stage 1: Understand the sampling design and data availability\\
    
    Stage 2: Conduct analyses with a range of models\\
    
    Stage 3: Evaluate and compare estimates \\
    
    Stage 4: Summarize, map, and visualize the prevalence estimates

    \vspace{.2in}

\noindent
The full pipeline are carried out using the \textit{surveyPrev} package \citep{surveyPrev} in R \citep{Rlang}. A reproducible report with all analysis scripts is included in the Supplementary Materials. 

\subsection{Stage 1: The sampling design and data availability} \label{sec:1-exploratory}
\begin{figure}[!ht]
    \centering
    \includegraphics[width=0.9\linewidth]{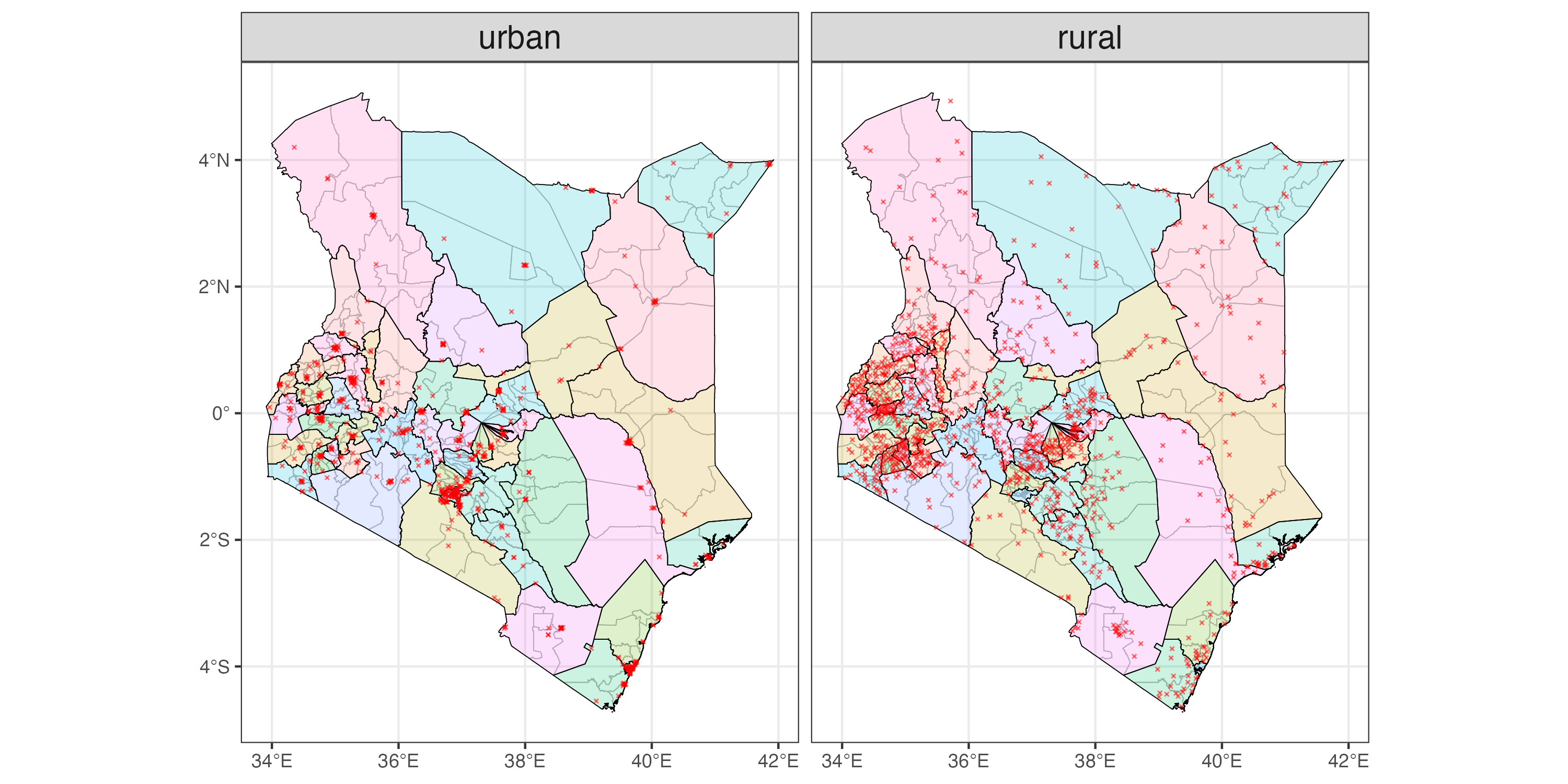}
    \caption{Map of the sampled clusters in the 2022 Kenya DHS by urban/rural status.}
    \label{fig:cluster-map}
\end{figure}

Understanding the survey data and how they are collected is critical in any modeling tasks, including prevalence mapping. Data availability has important implications for model choices. 
In our working example, we use the GPS coordinates of the clusters, which are available for most DHS surveys and some MICS surveys, and administrative boundaries from the GADM database \citep{gadm} to determine which area each cluster belongs to. Mapping coordinates to areas may create incorrect assignments for clusters due to jittering or imprecise boundary maps. \rev{Comparing the area assignments with spatial information recorded in the survey, e.g., Admin-1 labels, is recommended whenever such information exists. An example of the detailed steps in correcting potentially mis-assigned clusters is included in the Supplementary Materials.} In practice, we expect the potential impact to be minor unless the geography is very small. \rev{Further discussion of this issue can be found in \citet{altay2025impact}.}
Figure \ref{fig:cluster-map} shows the location of sampled clusters across Admin-1 and Admin-2 areas, by urban and rural strata. The median number of clusters is $35$ across the $47$ Admin-1 areas and $5$ across the $300$ Admin-2 areas. There are $6$ Admin-2 areas without a sampled cluster and $5$ with only one sampled cluster. The sample size of women who had a birth or stillbirth within the $5$ years preceding the survey has a median size of $218$ across Admin-1 areas and $29$ across Admin-2 areas. While it is difficult to provide a specific threshold for sample size, both the number of clusters and the individuals are reasonably large for Admin-1 areas but limited in most of the Admin-2 areas.  


\begin{figure}[htb]
    \centering
    \includegraphics[width=0.5\linewidth]{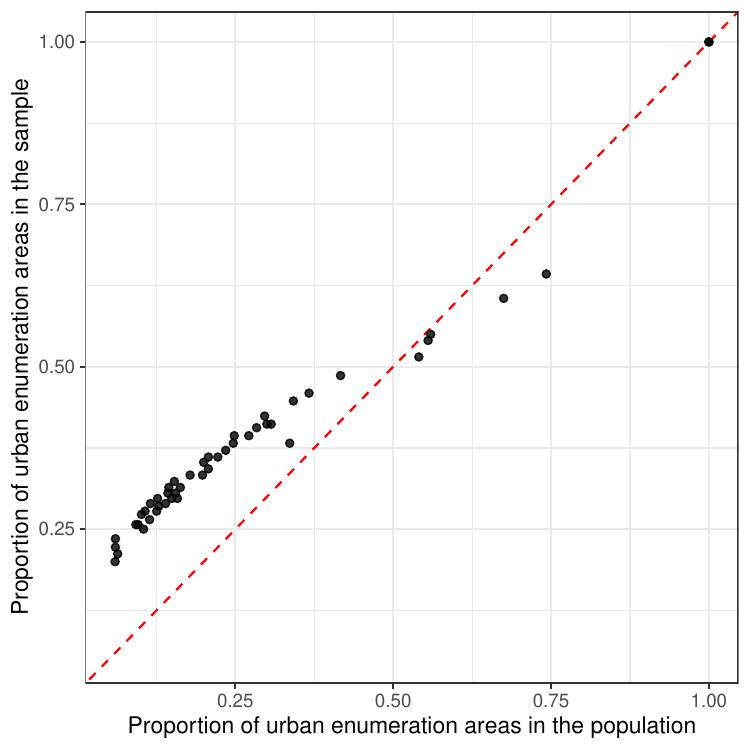}
    \caption{Comparison between the fraction of sampled clusters that are urban and the fraction of population urban clusters (enumeration areas) in each Admin-1 area.}
    \label{fig:over-sample}
\end{figure}

Another important consideration is the effect of within-area urban/rural stratification. Most of the household surveys in LMICs, including DHS and MICS, are stratified by urban/rural status within Admin-1 areas. 
The population proportions of urban/rural clusters are usually released in the survey report or are available from the census. Figure \ref{fig:over-sample} shows the sampled proportion of urban clusters in each Admin-1 area compared to population proportions. We observe that urban clusters are over-sampled in most of the areas. The coverage of ANC4+ is higher in urban areas (the odds ratio from a weighted logistic regression $1.55$). Therefore, models that do not use the weights and do not account for the stratification will overestimate the prevalence.
 The exact impact of the over-sampling of urban areas can only be determined by analyzing data under a specific analysis model, which we investigate further  in Section \ref{sec:unit}. 

\subsection{Stage 2: Models and analysis}

We propose three sets of models for routine analysis: direct estimation, area-level models, and cluster-level models. The overall approach we recommend is to fit all models, where possible, to assess sensitivity but to default to simpler models if they provide sufficiently precise estimates. In the rest of this section, we describe the three classes of models and discuss their advantages and limitations. 
\rev{We have omitted sensitivity analysis for some model specifications, such as prior choices, as the results in our example are not sensitive to these choices, but it is good practice to conduct more comprehensive sensitivity analysis in practice. }

To fix notation, consider a finite target population in a country with $N$ individuals living in $M$ areas (Admin-1 or Admin-2, depending on the target resolution). Let $y_j \in \{0, 1\}$ denote the outcome value for the $j$-th individual. We let $p_i$ denote the area-level prevalence for the $i$-th area. We let $U_i$ be the set of individuals in the target population in the $i$-th area with size $N_i = |U_i|$ and $S_i \in U_i$ denote the set of $n_i=|S_i|$ sampled individuals in the $i$-th area. Then the true prevalence in the $i$-th area is $p_i = (1/N_i) \sum_{j \in U_i} y_j$. For all sampled individuals, we observe the design weight $w_j$, which is the inverse probability of the $j$-th individual being included in the sample for $j \in S_1 \cup ... \cup S_M$.

\subsubsection{Direct estimation} \label{sec:direct}
Our starting point for prevalence analysis is direct estimation. The term `direct' refers to estimating the prevalence of an area using only the response data from that area \citep{rao:molina:15}. An example of a direct estimator is the design-based weighted estimator \citep{hajek:71},
\[
\hat p_i^{w} = \frac{\sum_{j \in S_i} w_j y_j}{\sum_{j \in S_i} w_j}, \;\;\;\; i = 1, ..., M.
\]
The variance of $\hat p_i$ can be easily estimated with standard software for survey data. The direct estimates and the associated uncertainty measure account for the sampling design via weighting, avoiding the bias due to informative sampling. \rev{When there are many clusters in the sample for a given area, confidence intervals constructed from the normality assumption will be accurate for the area.} Direct estimation is based on minimal assumptions since there is no explicit model for the data.
In practice, direct estimation often provides reasonable inference for Admin-1 areas, unless the outcome is very rare. The precision associated with the direct estimates, however, is usually comparatively small, since the estimates in each area only use data from that area alone. This can be contrasted with the area-level and unit-level approaches, which borrow strength from the totality of data across all areas, which effectively increases the sample size.

When data are sparse in an area, the uncertainty of the estimator can be too high to be usable. When there are a small number of clusters in an area, the point estimate of the prevalence may be $0$ or $1$, and in these cases, the usual formulas for calculating the variance equal zero. Other sparse data configurations may also produce unreasonably small estimates of uncertainty. In addition, even when point estimates and standard errors can be computed, the resultant uncertainty interval, based on the asymptotic normality assumption, can be inaccurate when the sample size is small. 
 
\begin{figure}[!ht]
    \centering
    \includegraphics[width=\linewidth]{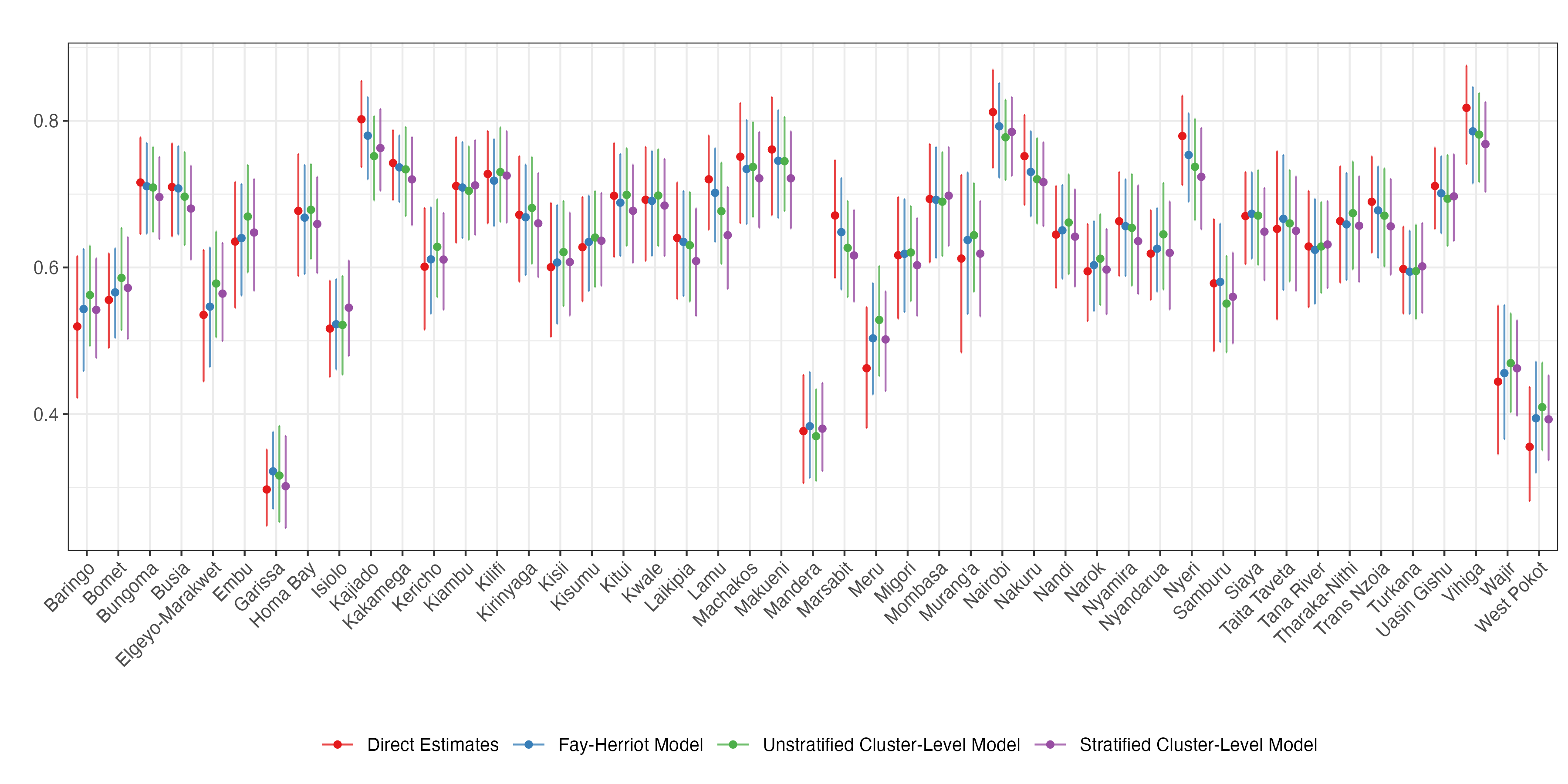}
    \caption{Point estimates and 95\% uncertainty intervals for ANC4+ coverage across Admin-1 areas in Kenya, estimated using different models. Covariates are not included in all models.}
    \label{fig:line-comparison}
\end{figure}

\begin{figure}[!ht]
    \centering
    \includegraphics[width=\linewidth]{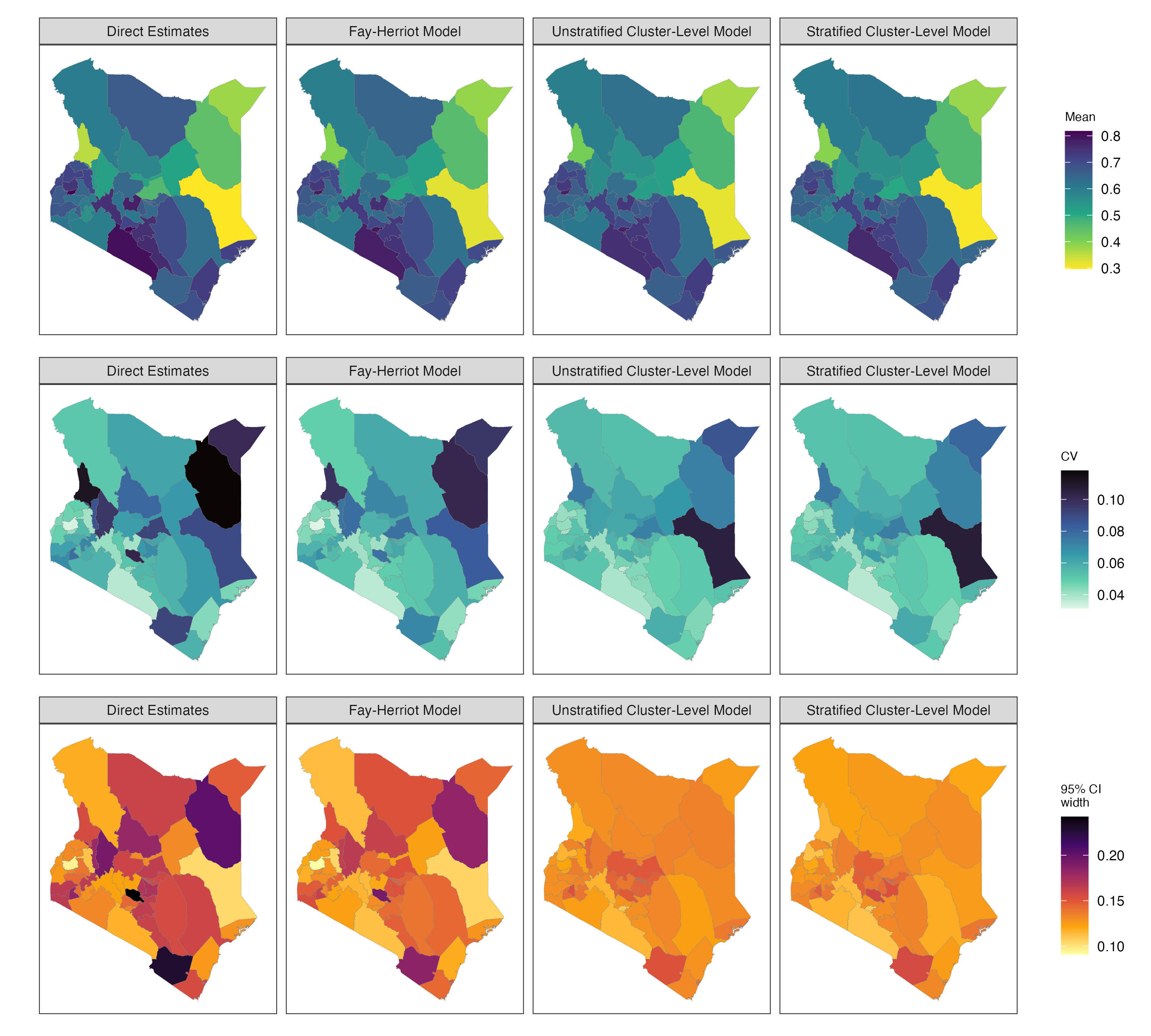}
    \caption{Maps of estimated prevalence (top row), coefficient of variation (middle row), and the width of 95\% uncertainty interval (bottom row) for ANC4+ coverage across Admin-1 areas in Kenya, estimated using different models. Covariates are not included in all models.}
    \label{fig:map-comparison1}
\end{figure}

\begin{figure}[!ht]
    \centering
    \includegraphics[width=\linewidth]{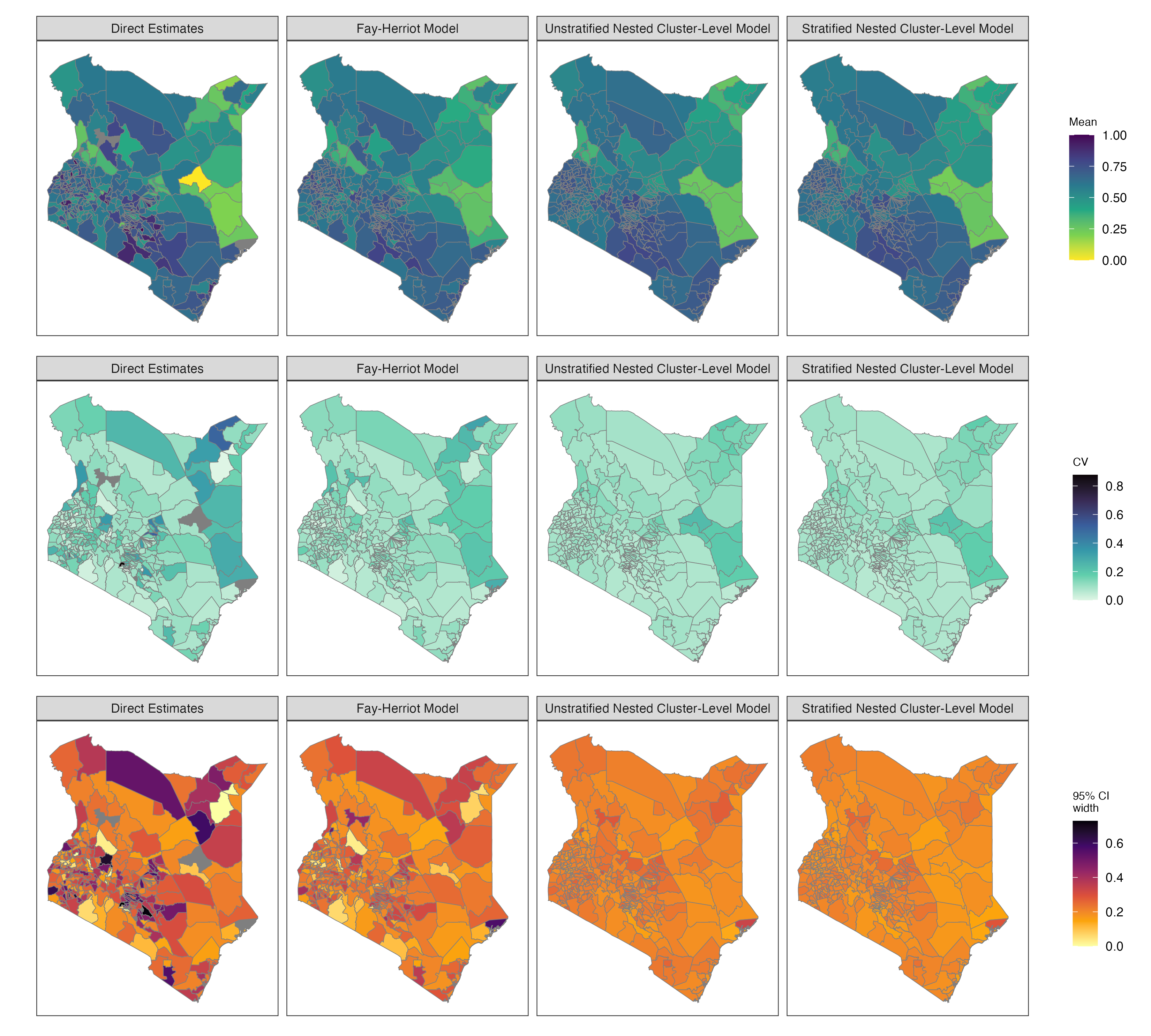}
    \caption{Maps of estimated prevalence (top row), coefficient of variation (middle row), and the width of 95\% uncertainty interval (bottom row) for ANC4+ coverage across Admin-2 areas in Kenya, estimated using different models. Covariates are not included in all models.}
    \label{fig:map-comparison2}
\end{figure}

Figure \ref{fig:line-comparison} shows the point estimates and 95\% uncertainty intervals of different models we consider in the workflow. The point estimates are mostly comparable across all models at Admin-1. Figure \ref{fig:map-comparison1} visually compares two commonly reported metrics of uncertainty: the coefficient of variation (CV), $\sqrt{var(\hat p_i)} / \hat p_i$, and the width of the $95\%$ uncertainty intervals. Both metrics are typically larger for the direct estimates compared to the smoothing models.
Figure \ref{fig:map-comparison2} shows the same comparison for Admin-2 estimates. The larger uncertainty of the direct estimates in finer resolutions can be observed more clearly. \rev{The widest 95\% uncertainty interval for Admin-1 direct estimates is $0.242$, and more than $60\%$ of Admin-2 areas have 95\% uncertainty intervals that surpass this.}

\subsubsection{Area-level models} \label{sec:area}

Area-level models, or Fay-Herriot models \citep{fay:herriot:79}, are the most common SAE method and have seen widespread use in official statistics. Under this approach, the direct estimates are linked together via a hierarchical model. For prevalence modeling, we first transform the direct estimates to the entire real line via $\hat \theta_i^{w} = \logit(\hat p_i^{w})$, where $\logit(p) = \log(p / (1 - p))$, and derive the associated asymptotic variance estimate of $\hat \theta^{w}$ using the delta method, i.e., $\hat V_i = \widehat{var}(\hat p_i^{w}) / [\hat p_i^{w}(1-\hat p_i^{w})]^2$. The Fay-Herriot model assumes
\begin{eqnarray}
\hat \theta_i^{w} \mid \theta_i &\sim& N(\theta_i, \hat V_i) \label{eq:FH1} \\
\theta_i &=& \alpha + \bx_i^T\bbeta + u_i, \;\;\;\;i = 1, ..., M, \label{eq:FH2}
\end{eqnarray}
where $\alpha$ is the intercept, $\bx_i$ are area-level covariates and $\bbeta$ are the associated coefficients. \rev{Equation (\ref{eq:FH1}) is referred to as the {\it sampling model} and equation (\ref{eq:FH2}) the {\it linking model}.} The latent prevalence at the $i$-th area is then $\expit(\theta_i)$, where $\expit(\theta) = \exp(\theta)/(1+\exp(\theta))$. The original Fay-Herriot model assumed that the random effects $u_i$ are independent and identically distributed (IID) from a normal distribution. As discussed before, we usually expect residuals to remain spatially correlated when covariates are limited. \rev{A wide range of spatial random effect models have been proposed and compared in the literature \citep[see e.g., ][]{chung:datta:20, chandra2019small}.} One approach that we recommend as a default choice is a model that decomposes the random effect into the sum of an unstructured IID normal random effect and a spatially structured intrinsic conditional autoregressive (ICAR) random effect, which is known as the BYM model \citep{besag:york:mollie:91}. In our analysis, we adopt the reparameterization known as the BYM2 model \citep{riebler:etal:16}, 
\[
\bm u = \sigma (\sqrt{1-\phi}\bf{e} + \sqrt{\phi} \bm{S}),
\]
where $\sigma$ is the total standard deviation, $\phi$ is the proportion of the variance that is spatial, $\bf{e}$ is IID standard normal random variable and $\bm S$ follows a scaled ICAR prior so that the geometric mean of the marginal variances of $S_i$ is equal to $1$. We adopt the penalised complexity (PC) priors for the hyperparameters \citep{simpson:etal:17}. \rev{The PC priors provide robust shrinkage of model components to simpler base models and have been used successfully in various contexts \citep[see e.g.,][]{sorbye2017penalised,fuglstad:etal:19b}. Compared to the standard conjugate priors, the PC priors are specified by more intuitive probability statements.} 
In our working example, for the total standard deviation, we specify $\mbox{Prob}(\sigma > 1) = 0.01$ and for the proportion of the variation that is spatial, we have $\mbox{Prob}(\phi > 0.5) = 2/3$. 
Posterior computation is performed with the integrated nested Laplace approximation (INLA) approach \citep{rue:etal:09}. INLA is extremely popular for carrying out Bayesian inference for dependent data in general and spatial data in particular \citep{blangiardo:cameletti:15,krainski:etal:18,osgood-zimmerman:wakefield:19} because it is accurate and very fast. \rev{Under the Bayesian framework, inference is conducted by evaluating the posterior distributions and credible intervals of the model parameters. In our context, the parameter of interest is the latent prevalence, $\expit(\btheta)$.}

The rationale for the Fay-Herriot model is that the similarity of prevalences across the study areas may be leveraged to fine-tune the estimates in each area. Since data from all areas are used for estimation in each area, this is an example of an `indirect' estimate. \rev{The model accounts for the sampling design by using the design-based estimates, $\hat p_i^w$, and their variances, $\widehat{var}(\hat p_i^w)$ as the input. The latter goes to zero as the sampling fraction in an area goes to 1, and the weighted estimate equals the population prevalence in the area so that} the estimates produced by the model are design consistent, which is very appealing. By collectively modeling all the data, uncertainty in each area is reduced on average. The model that links the areas together introduces additional modeling assumption, though this is typically not strong. We expect on average the mean squared errors of the estimates to be reduced compared to direct estimation, as the gain in precision will offset the bias due to shrinkage. The interval estimates based on hierarchical models, however, do not typically have the usual frequentist coverage when each area is viewed in isolation. Across the map the coverage will be close to nominal \citep{burris2020exact}, but interpreting the uncertainty in each area is more tricky.

A key disadvantage of Fay-Herriot models is that they are based on direct estimates. When data are extremely sparse, direct estimates or their variance estimates are unavailable or unstable, which renders the Fay-Herriot model unusable or unreliable. \rev{To alleviate variance instability, the sampling variances may be modeled using generalized variance functions, which leverage the mean-variance relationship between $\theta_i$ and $V_i$, as well as covariates \citep{otto:bell:95, mohadjer_hierarchical_2012, franco_applying_2013, liu_hierarchical_2014,gao2023spatial}. 
When fitting the Admin-2 level Fay-Herriot model, the design-based variance formula does not produce an estimate, or produces an estimate that is zero or close to zero for  $12$ out of $300$ areas. For these Admin-2 areas, we supplement the data with a ``phantom'' cluster that has prevalence equal to the prevalence in the Admin-1 area within which the Admin-2 area is contained, and with the sum of the weights equal to the average sum across all observed clusters in the Admin-1 area. For further details, see \cite{wakefield:etal:25}.}

 Figures \ref{fig:line-comparison} to \ref{fig:map-comparison2} present the results of the Fay-Herriot model where no covariates are included
 We observe that the Fay-Herriot model estimates are, in general, shrunk from the direct estimates, though to a lesser extent compared to the cluster-level models that we describe next.

\subsubsection{Cluster-level models} \label{sec:unit}

SAE models dealing with individual outcomes directly are known as unit-level models. In our context, modeling individual responses (the `units') using a Bernoulli likelihood is equivalent to a cluster-level binomial model. More specifically, for a survey consisting of $C$ clusters, we let $c[j]$ be the cluster index of the $j$-th individual. \rev{When there is no ambiguity of resolution, we use $i[c]$ to denote the index of the area that the $c$-th cluster resides in. For nested models where both Admin-1 and Admin-2 level components are included, we use $a[c]$ and $a[i]$ to denote the corresponding Admin-1 index for the $c$-th cluster or the $i$-th Admin-2 area.} Let $Y_{c}$ and $n_c$ denote the number of positive outcomes and the number of sampled individuals from the $c$-th cluster, respectively, i.e., $Y_{c} = \sum_{j: c[j] = c} y_{j}$ and $n_{c} = \sum_{j: c[j] = c} 1$. We consider a cluster-level model with a beta-binomial likelihood,
\begin{align*}
Y_c \mid p_c &\sim \mbox{BetaBinomial}(n_c, p_{c}, d) .
\end{align*}
In this formulation, $p_c$ is the latent prevalence for the $c$-th cluster, and $d$ is the overdispersion parameter that captures the additional within-cluster variation. 
This class of beta-binomial models has been used in modeling a variety of indicators in the recent literature. A detailed review of different binomial models was carried out in \citet{dong:wakefield:21}. 

To link the cluster-level prevalence to area-level prevalence, \rev{a simple unstratified model is
\[
\logit(p_{c}) = \alpha + \bx_{c}^T\bbeta + u_{i[c]}.
\]
The intercept and spatial random effects are modeled in the same way as in the Fay-Herriot model, and $\bx_c$ are cluster-level covariates. The covariates may also be at the area level, which simplifies the aggregation step discussed in Section \ref{sec:area-covariate}.  When modeling at Admin-2 level, it is often also reasonable to include a separate intercept for each Admin-1 level to reduce the shrinkage across Admin-1 areas. This leads to a nested unstratified model,
\[
\logit(p_{c}) = \alpha_{a[c]} + \bx_{c}^T\bbeta + u_{i[c]}.
\]
The stratification over Admin-1 areas is either implicitly accounted for by the inclusion of random effects, $u_{i[c]}$ or explicitly by $\alpha_{a[c]}$. We use the term `unstratified' to refer to the fact that the model does not account for any within-area stratification, as discussed in Section \ref{sec:1-exploratory}. To account for the additional stratification by urbanicity, we consider the following three stratified models:
\begin{align*}
\mbox{(non-nested)} \;\;\;\; 
\logit(p_{c}) &= \alpha + \gamma \bm 1_{c \in \mbox{\footnotesize{rural}}} +  \bx_{c}^T\bbeta + u_{i[c]},\\
\mbox{(nested)} \;\;\;\; 
\logit(p_{c}) &= \alpha_{a[c]} + \gamma \bm 1_{c \in \mbox{\footnotesize{rural}}}+ \bx_{c}^T\bbeta + u_{i[c]},\\
\mbox{(nested with interaction)} \;\;\;\; 
\logit(p_{c}) &= \alpha_{a[c]} + \gamma_{a[c]} \bm 1_{c \in \mbox{\footnotesize{rural}}} + \delta_{a[c]} +  \bx_{c}^T\bbeta + u_{i[c]} .
\end{align*}
}
\rev{All three models account for the within-area stratification by including terms corresponding to the urban/rural status. The first two models assume the urban/rural effect, $\gamma$, is constant across all areas. The third model further assumes that the urban/rural associations vary over Admin-1 areas. 
Selecting which model to use is context-specific. In our example, a likelihood ratio test \citep{lumley2014tests} for the interaction terms is included in the Supplementary Materials, together with further model diagnostics of all three models in terms of the WAIC scores. Both procedures favor the nested model over the other two. There is not enough evidence for spatially varying urban/rural associations. Therefore, we focus on the nested model in the rest of the paper.}

It is worth noting that most of the geostatistical models in the literature ignore the stratification variable. \citet{giorgi2021model} argued that the sampling design could be implicitly accounted for by including adequate covariates. 
In practice, however, it is difficult to assess whether a specific set of covariates is sufficient to adjust for the sampling design. Thus, we view the inclusion of the stratification variable as a necessary step, as long as the prevalences are associated with the urban/rural designation and there is evidence of over-sampling of urban or rural clusters.

The cluster-level models can deal with very sparse data, as they do not rely on the asymptotic properties of the direct estimates. \rev{The existence of data idiosyncrasies, such as zero response counts, can be easily accommodated, in contrast with weighted estimators.} If the parametric likelihood model is a reasonable approximation to the true underlying data generating mechanism, this approach is an efficient use of the data. Similar to the Fay-Herriot model, the cluster-level models also introduce bias, due to shrinkage, but reduce uncertainty. Hence, the mean squared errors of the estimates will be reduced on average. 
On the other hand, a major limitation of the cluster-level models is the assumption of an individual-level sampling model, which cannot be easily checked. \citet{dong:wakefield:21} provided a more in-depth discussion of different sampling models for binary outcomes, and the beta-binomial model we consider here is among the best-performing candidates in their empirical comparison. Also, cluster-level models are more computationally intensive to fit, though this is typically not an issue with a scalable implementation using INLA. \rev{Last, and not least, additional post-processing steps are necessary to aggregate cluster-level predictions into area-level estimates, and external information is often needed, which we discuss in the next two subsections.}

\subsubsection{Cluster-level aggregation with area-level covariates}\label{sec:area-covariate}
\rev{The inclusion of the urban/rural effect and cluster-level covariates comes at a cost when producing area-level estimates. We first consider the case with area-level covariates $x_i$}. For the unstratified model, the prevalence of the $i$-th area is simply $\theta_i = \expit(\alpha + \bx_i^T\bbeta + u_i)$. For the stratified model, however, an additional aggregation step is needed. As an example, for the nested model,
\[
\theta_i = (1 - q_i) \times \expit(\alpha_{a[i]} + \gamma + \bx_i^T\bbeta + u_i) + q_i\times \expit(\alpha_{a[i]} + \bx_i^T\bbeta + u_i),
\]
where $q_i$ is the proportion of the population in  area $i$ that is urban. The population parameter $q_i$ usually cannot be reliably estimated from the survey at fine resolutions. The interpretation of $q_i$ is also subtle. The urban/rural designation in household surveys is usually based on a previous census classification. The actual status of a cluster can change over time. Thus, $q_i$ represents the fraction of the target population who, at the time of the survey, reside in areas that were classified as urban in the last census. 

Two challenges exist in obtaining these fractions. First, the complete list of clusters and their urban/rural designation is never released to the public. The proportion of urban population is sometimes reported, but usually at the Admin-1 or national level only. 
Second, the target population can vary significantly across indicators, and the total population of a country may not be a good proxy for it. For example, when estimating ANC4+ coverage, the target population consists of women who had a recent birth or stillbirth, while for estimating the neonatal mortality rate, the target population consists of all births within the relevant time period. Therefore, these aggregation fractions need to be carefully specified and estimated from external data sources. 

Reconstructing $q_i$ forms a line of research on its own. One recent proposal using classification models to reconstruct urban/rural partitions can be found in \citet{wu2024modelling}. Here we adopt a simple thresholding approach that does not require sophisticated model tuning or high-quality external data. The approach relies on the simplifying assumption that the urban locations have higher population density than the rural locations. The process consists of the following steps: 
\begin{enumerate}
    \item Identify the year when the sampling frame was defined (usually the last census). Denote this year as $T_0$, and the year the survey took place as $T$. Obtain the total population raster for the country at year $T_0$ from sources such as WorldPop \citep{tatem2017worldpop}.
    \item For each Admin-1 area, use the population raster to identify a threshold such that the proportion of the population living in pixels with higher population density matches the reported fraction of urban population in this area at $T_0$. These reported fractions can usually be found in the survey report. The pixels with population above the corresponding threshold form the spatial partition of urban areas. 
    \item Identify the age-sex-specific sub-population with the closest match to the target population at year $T$. For example, for ANC4+, we use the female population of age $15$-$49$ as the proxy target population from the 2022 gridded population estimates for Kenya \citep{kenyapop22}. 
    \item Use the pixel-level population map of this sub-population at year $T$ and the urban partition to compute the fraction of the target population residing in the area designated as urban at year $T$.
\end{enumerate}

\begin{figure}[!ht]
    \centering
    \includegraphics[width=\linewidth]{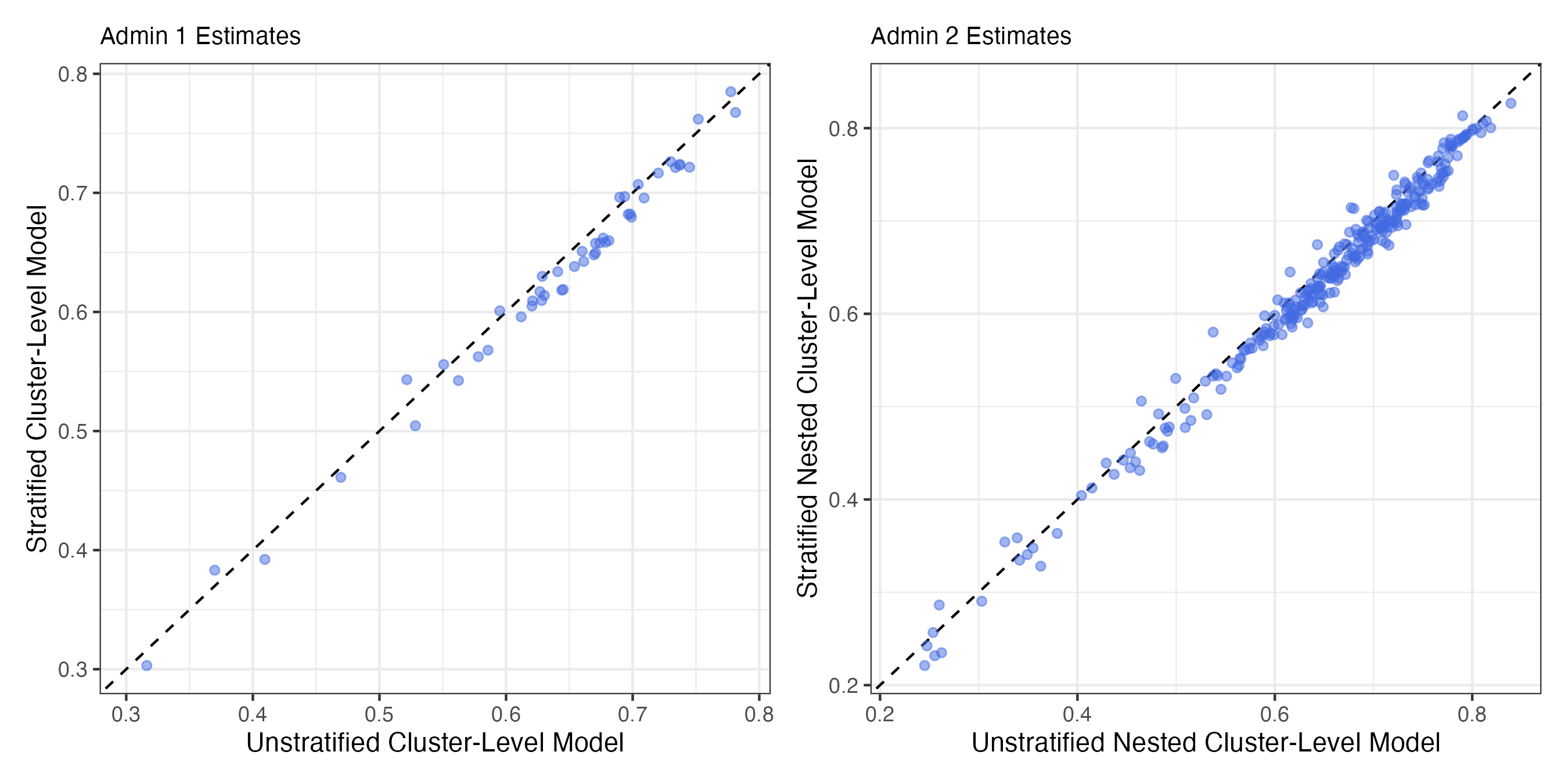}
    \caption{Comparing the estimated prevalence using the unstratified and stratified model at Admin-1 level (left) and Admin-2 level (right). Covariates are not included in the models.}
    \label{fig:ur-scatter}
\end{figure}

Given the complexity of constructing the aggregation weights, in practice, we recommend first fitting both the unstratified and stratified models without aggregation. The urban- and rural-specific estimates from the stratified model allow one to gauge how sensitive the aggregated results are regarding the unknown urban fraction. If the log odds ratio associated with urban/rural, $\gamma$, is estimated to be very close to $0$, the aggregated results will be very similar to the unstratified model regardless of the aggregation weights. In our analysis, $\gamma$ is estimated to be $-0.453$ (95\% credible interval: $[-0.564, -0.343]$), which indicates lower coverage of ANC4+ in rural areas on average. 
Given the over-sampling of urban clusters discussed in Section \ref{sec:1-exploratory}, and the significant association between ANC4+ and urbanicity of clusters, the unstratified models are likely to be biased and produce higher estimates. This is confirmed in Figure \ref{fig:ur-scatter}. Therefore, the stratified model is preferred.

Figure \ref{fig:map-comparison2} shows the estimates produced by the nested cluster-level models without covariates. \rev{Compared to the Fay-Herriot model, Admin-2 level estimates show greater shrinkage towards the Admin-1 average.  This is expected when data are sparse and do not contain sufficient information for the model to deviate from being `flat' within each Admin-1 area. Without the Admin-1 fixed effect, the estimates are more likely to be shrunk towards the overall mean, which produces a map that is too flat.}
 
\subsubsection{Cluster-level aggregation with cluster-level covariates}\label{sec:cluster-covariate}

\rev{
When cluster-level models are used, the aggregation step becomes more complicated, as predictions must be made for both the sampled and unsampled clusters. Take the nested stratified model, for example, 
\[
\theta_i = \sum_{c: i[c] = i} q'_c \times  \expit(\alpha_{a[c]} + \gamma \bm 1_{c \in \mbox{\footnotesize{rural}}} + \bx_c^T\bbeta + u_i),
\]
where $q'_c$ is the proportion of population in the $c$-th cluster within $i$-th area. Notice that the summation is over all clusters in the population, rather than the sampled clusters. In practice, the population frame is almost always unknown to researchers, thus the summation over clusters are usually approximated by a summation over a fixed set of grids where population and covariate values are known. 
 Compared to the models with only area-level covariates in Section \ref{sec:area-covariate}, stronger modeling assumptions are needed here as the aggregation step requires classifying each pixel as urban or rural, instead of just estimating the urban population proportion. We follow the same thresholding procedure in Section \ref{sec:area-covariate} to assign urban/rural status to each pixel.
}
 
\subsubsection{Inclusion of covariates} \label{sec:covariate}

In general, covariates strongly related to the indicator of interest can usually improve prevalence estimates. \rev{Cluster-level covariates are often more informative than area-level covariates. However, with non-linear models, aggregation requires them to be known for all clusters in the population. In practice, this requirement is rarely attained and instead, we require the covariates to be available on a grid, which approximates the cluster-level availability but leads to a further approximation step when aggregation from cluster to area is carried out.  We also focus on improving prevalence estimates rather than interpreting the association between covariates and the outcome. Therefore, while some covariates may not be causally related to the outcome of interest, we may still choose to include them in the model if they are correlated with the outcome. Given the common scarcity of available covariates, we omit the discussion of variable selection in this paper. Often, this step will be carried out in a more informal manner and will be strongly based on the availability of well-measured covariates. We have a preference for avoiding covariates that are the subject of excessive modeling before they are used. Readers who wish to examine more formal procedures are referred to procedures in the literature, for example, \cite{hoeting2006model}, \cite{molina2019small} and \cite{michal2023small}. } 

\begin{figure}[!ht]
\includegraphics[width = \textwidth]{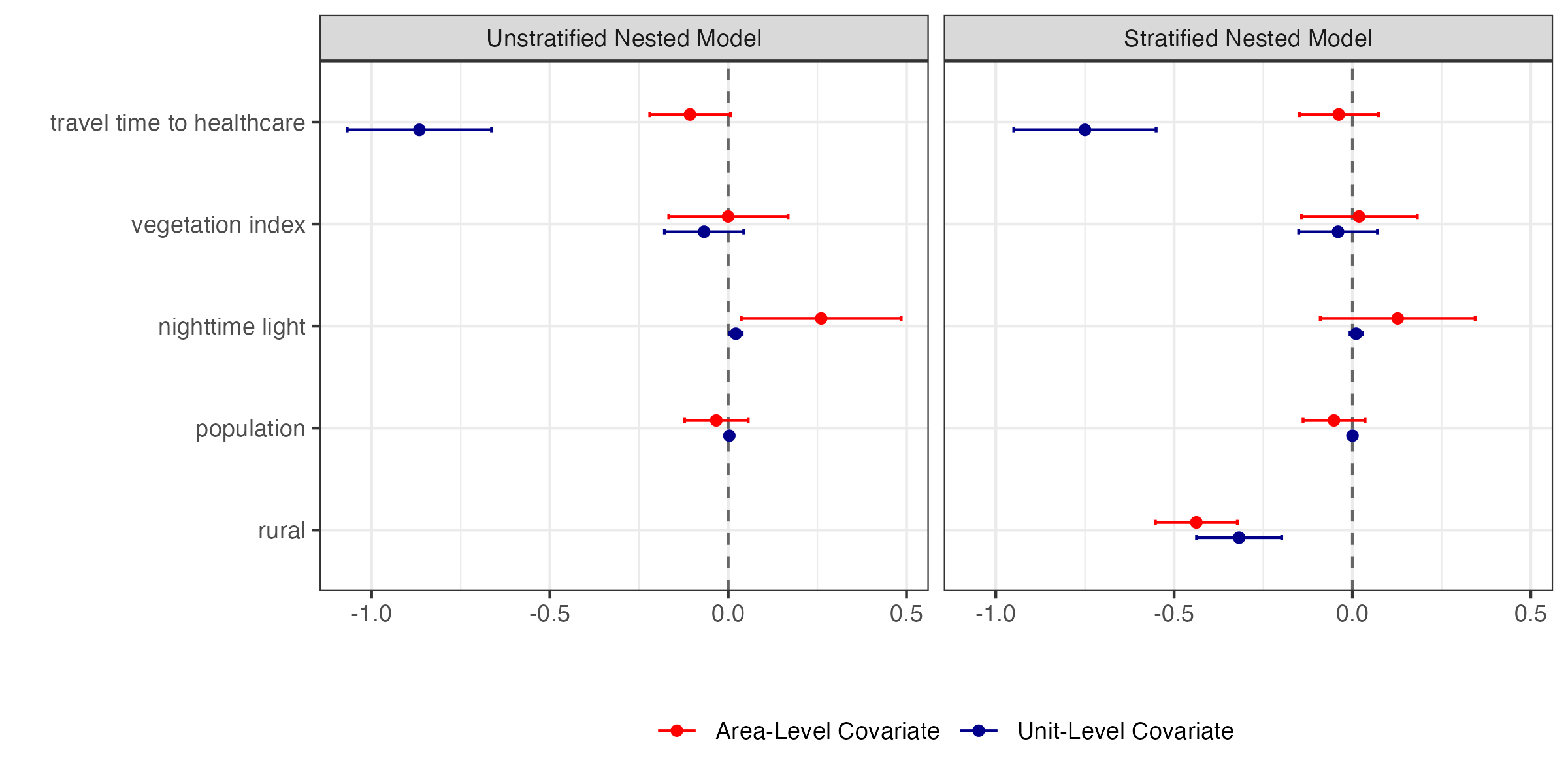}
\caption{Comparing the estimated mean and 95\% credible intervals of the coefficients of the fixed effects in the three models at both Admin-1 and Admin-2 levels.}
\label{fig:coefficient}
\end{figure}

\begin{figure}[!ht]
    \centering
    \includegraphics[width=\linewidth]{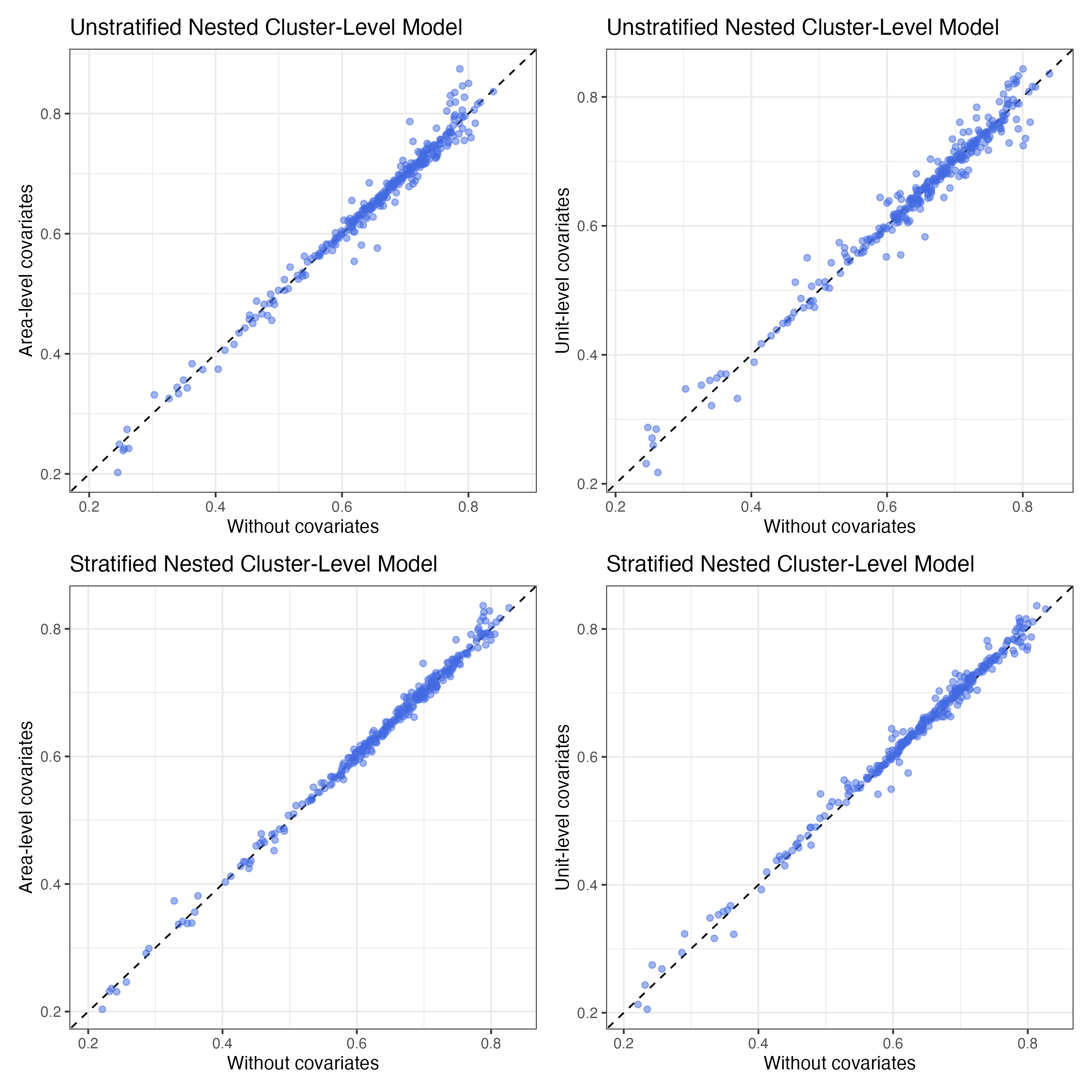}
    \caption{Comparing Admin-2 level model with and without area-level covariates (left) and unit-level covariates (right) in terms of the estimated prevalence for the nested models that do not account for urban/rural stratification (top row) and nested and stratified model (bottom row). }
    \label{fig:cov-scatter}
\end{figure}

 \rev{We consider four covariates as an illustrative example}: the total population \citep{tatem2017worldpop}, night time lights \citep{roman2018nasa}, vegetation index \citep{didan2015modis}, and travel time to the nearest healthcare facility \citep{weiss2020global}. Both nighttime lights and vegetation index are averaged over the period of 2018--2022. \rev{We consider both the pixel-level covariates and area-level covariates by aggregating them into Admin-1 or Admin-2 level averages, weighted by the population at each pixel}. The covariates are standardized to have mean $0$ and unit variance. \rev{Figure \ref{fig:coefficient} shows the estimated regression coefficients for Admin-2 level cluster-level models using either area- or pixel-level covariates.} The signs of the associations between each covariate and the outcome are consistent among models. Comparing the two cluster-level models, the effects of covariates are slightly attenuated after accounting for the urban/rural designation of the clusters. \rev{Figure \ref{fig:cov-scatter} shows that the final Admin-2 level estimates do not change much with the inclusion of covariates,
 which has also been found in other work (see for example, \cite{wakefield:etal:19}). 
 However, we would rather explain between-area differences in terms of covariates, and we acknowledge that further work, both methodological and through practical examples, is required with respect to covariate modeling.}

\subsection{Stage 3: Model evaluation and estimate comparison} \label{sec:3-eval}

Evaluating and comparing estimates from different models is an important step in producing small area statistics. It is often helpful to first rule out models that are inappropriate to use. While not exhaustive, we find the following steps often provide a useful path towards this goal. 

\begin{enumerate}
\item \rev{Are there sufficient data?} Areas without samples cannot produce a direct estimate. Thus, if too many areas have no data or only sparse data, direct estimates will contain many missing values. Consequently, estimates of Fay-Herriot models are also less trustworthy, as they rely heavily on random effects and covariates to impute the prevalence of those areas. 
This is more likely to occur when modeling at fine resolutions, and in this scenario, cluster-level models may be the only viable approach. It may also be reasonable to consider adjusting the target resolution of the estimates if data are highly sparse, since one must accept that below a certain level of data availability, no model can resuscitate the analysis.

\item \rev{Are estimates consistent with other models when aggregated to the national and the Admin-1 level?} When data are abundant, different methods should lead to similar estimates. For example, if there exist systematic differences between the national estimates, the smoothing model needs to be examined closely. As examples, the difference may be due to incorrect model implementation or specification, inappropriate covariates or external information (e.g.,~in aggregation of a unit-level model), or informative sampling not being properly accounted for. This diagnostic, however, is not sufficient to conclude a model is reliable. Models with systematic bias may still provide consistent aggregated estimates if the area-level biases cancel out. 

\item \rev{Are the estimates over-smoothed? Over-smoothing usually happens when observed data is sparse or when the model is misspecified. Severe over-smoothing can usually be identified by comparing the model-based estimates against another model or direct estimates through scatter plots. \rev{Figure \ref{fig:shrinkage} compares the point estimates of three smoothing models against the direct estimates. Usually, we expect the shrinkage of smoothed estimates to lead to the point cloud showing a tilted (attenuated towards the national prevalence) pattern, with more shrinkage in models of higher resolution.} Another useful check of over-smoothing is by comparing the variation in estimates against a more reliable model at a coarser level. Consider a set of prevalence estimates, $\theta_i$, for Admin-2 areas, we can calculate the posterior distribution of $v = \frac{1}{M-1}\sum_{i=1}^M(\theta_i - \bar{\theta})^2$. We then compare this with the same Admin-2 summary statistic computed from an Admin-1 model, where the prevalences in all Admin-2 areas nested within the same Admin-1 area are assumed to be constant. If the Admin-2 model has significantly smaller variation, over-smoothing is likely present. In our example, if we populate Admin-2 prevalence with the Admin-1 Fay-Herriot model, the estimated $v$ is $0.0122$ ($95\%$ credible interval: $[0.0096, 0.0147]$). For the stratified nested model at Admin-2, $v$ is estimated to be $0.0161$ ($95\%$ credible interval: [$0.0136, 0.0187]$), indicating the Admin-2 model indeed produces more variation than Admin-1 alternatives.}

\begin{figure}[!ht]
    \centering
    \includegraphics[width=\linewidth]{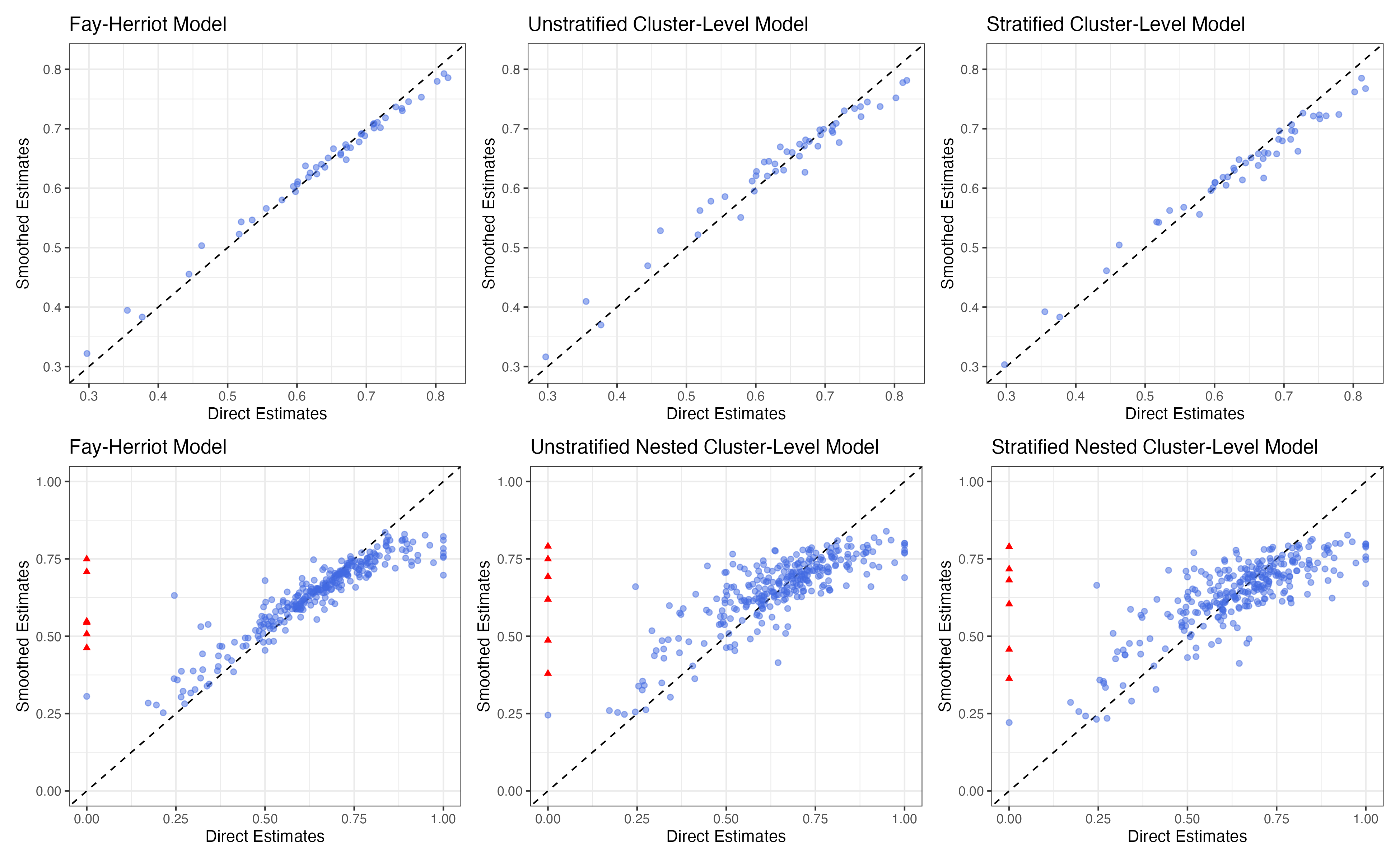}
    \caption{Comparing point estimates of the three smoothed models against the direct estimates. Top row: Admin-1 estimates. Bottom row: Admin-2 estimates. The red triangles correspond to areas where direct estimates cannot be computed.}
    \label{fig:shrinkage}
\end{figure}

\item \rev{Are the uncertainties too large to be useful? Government agencies often require certain precision when reporting estimates. For example, Statistics Canada has guidelines for area-level estimates, specified with respect to the coefficient of variation (CV) \citep{cloutier2014aboriginal}. Areas with CV less than $16.7\%$ can be used without restriction. Other metrics, e.g., width of uncertainty intervals, may also be used, and the choice of threshold for any metric is context-specific. For Admin-1 modeling in our example, direct estimates are probably good enough with reasonable uncertainty intervals and CVs, as can be seen in Figure \ref{fig:map-comparison1}. Further modeling may not provide much gain in practice. For Admin-2 modeling, as can be observed in Figure \ref{fig:map-comparison2}, smoothed estimates are usually necessary to obtain the desired precision. }

\item \rev{Are model parameters reasonably estimated? It is a good practice to visualize and check the model components of a smoothing model to confirm that the model is free from numerical or computational issues. Figure \ref{fig:random-effects} shows the estimated Admin-1 intercepts and Admin-2 random effects for the nested stratified cluster-level model. The Admin-2 random effects have much smaller magnitude than the Admin-1 fixed effects, as expected due to data sparsity.}

\begin{figure}[!ht]
    \centering
    \includegraphics[width=\linewidth]{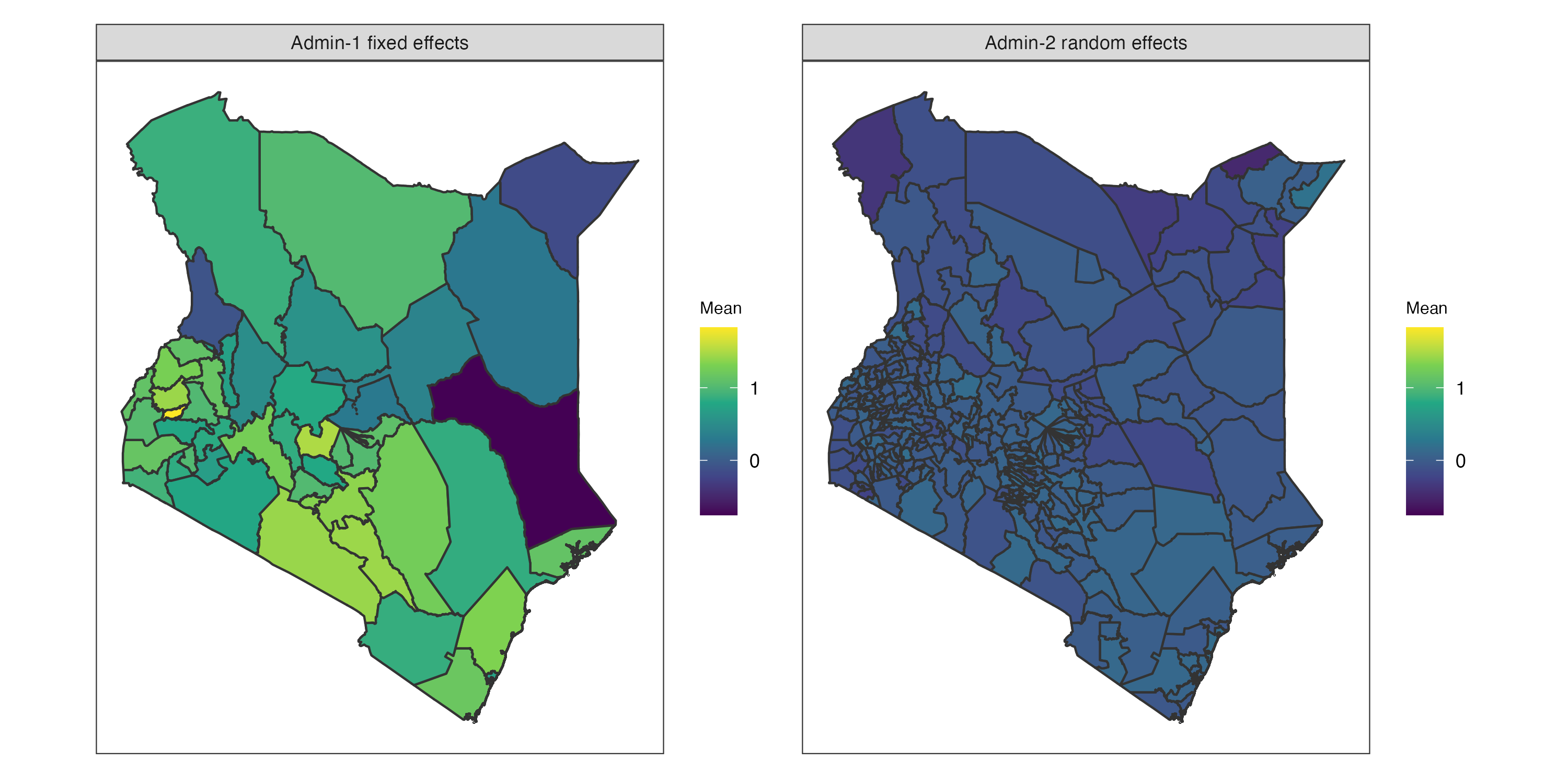}
    \caption{Posterior mean of the Admin-1 fixed effects and Admin-2 random effects in the nested stratified model without covariates.}
    \label{fig:random-effects}
\end{figure}

\item \rev{Which final model to report? When multiple models pass these previous checks, it is often of interest to pick a final model to report.
General guidelines for model comparison metrics are far from well-established and are an open area of ongoing research in the SAE literature. The gold-standard of model comparison is to compare the estimates with external `truth', usually the census \citep{merfeld2022combining, merfeld2024small}, but for many indicators in LMICs, such information is rarely available. Therefore, model performance is commonly assessed with simulations that are either model-based, i.e., samples are generated from a pre-specified true model, or design-based, i.e., re-sampled from a fixed population \citep[see e.g.,][]{molina2010small, torabi2008small}.} Different methods can be used to estimate the prevalence on the replicate samples and be compared against the truth \citep{tzavidis:etal:18}. However, population data are also rarely available to researchers outside of the national statistical agencies. 
Another common practice is leaving some areas out of the model fitting process and evaluating the out-of-sample predictive performance on those areas \citep{pfefferman:13}. 
\rev{\citet{merfeld2024small} argues that cross validation is usually the second-best option after census validation, though the best practice to implement cross validation is unclear in the literature.} 

\end{enumerate}

\subsection{Stage 4: Summarization and visualization} \label{sec:4-result} 

\begin{figure}[!ht]
    \centering
    \includegraphics[width=\linewidth]{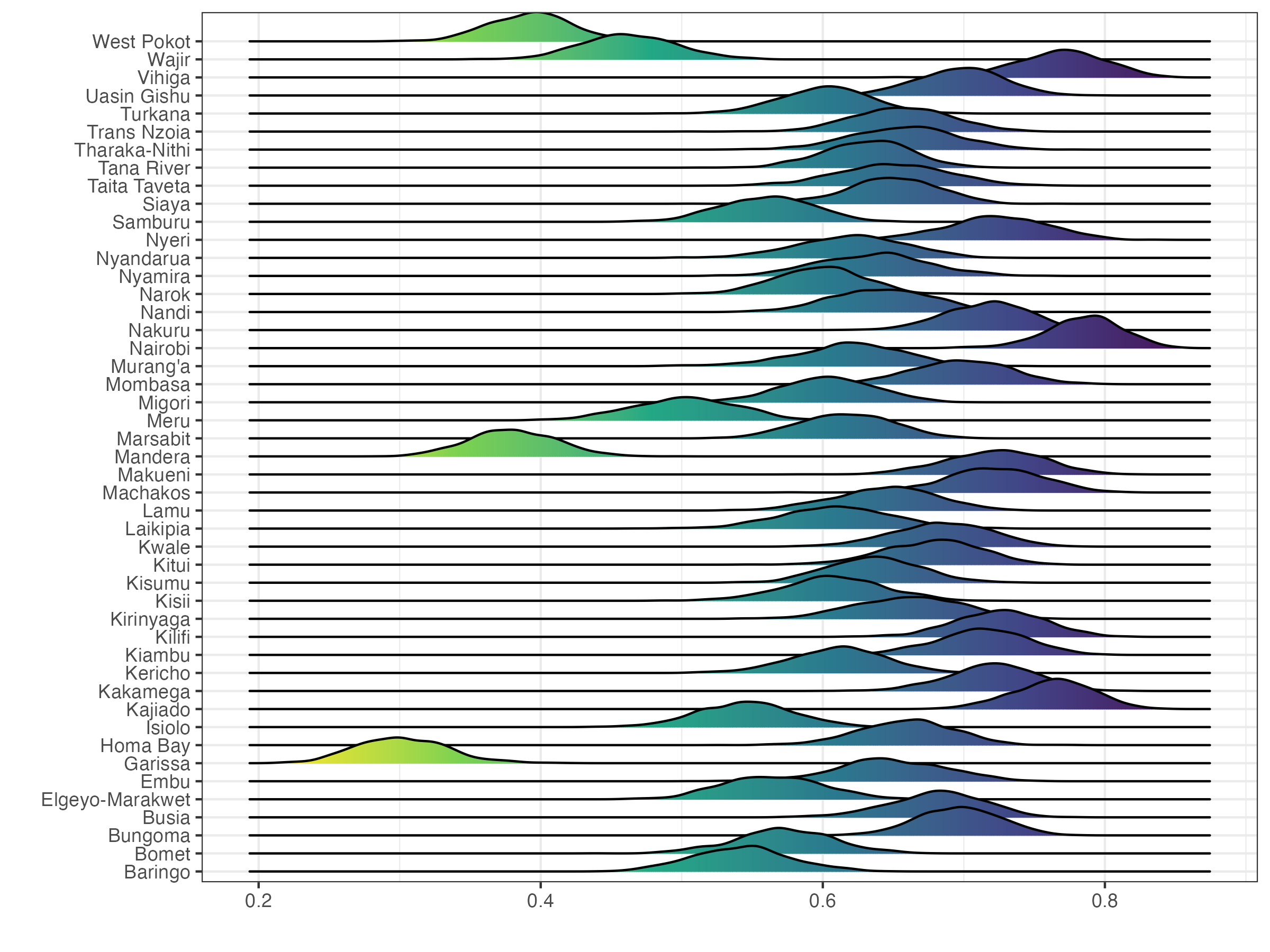}
    \caption{Ridge plot of the posterior distributions of the Admin-1 prevalence, based on the stratified cluster-level model.}
    \label{fig:ridge}
\end{figure}

In addition to numerical summaries, it is essential to develop visualization pipelines that enable practitioners to easily interpret the large number of model outputs. The interval plot in Figure \ref{fig:line-comparison} and choropleth map in Figures \ref{fig:map-comparison1} and \ref{fig:map-comparison2} are the most natural tools to examine the estimates and their uncertainties. The ridge plot in Figure \ref{fig:ridge}  provides another useful and more detailed view of the posterior marginal distributions of the estimates. A common theme behind these graphical summaries is the high uncertainty of the prevalence estimates. This is unavoidable, given limited data, but it has important implications for the interpretation of the results. For example, Figure \ref{fig:ridge} shows a large number of counties with similar distributions of ANC4+ coverage, between $50\%$ and $70\%$. Naively ranking these areas by their point estimates while ignoring such uncertainty could produce misleading results.
As the number of areas in consideration becomes larger, all of these plots can be difficult to examine visualy. Web-based interactive visualizations can be very useful. 

\begin{figure}[!ht]
    \centering
    \includegraphics[width=\linewidth]{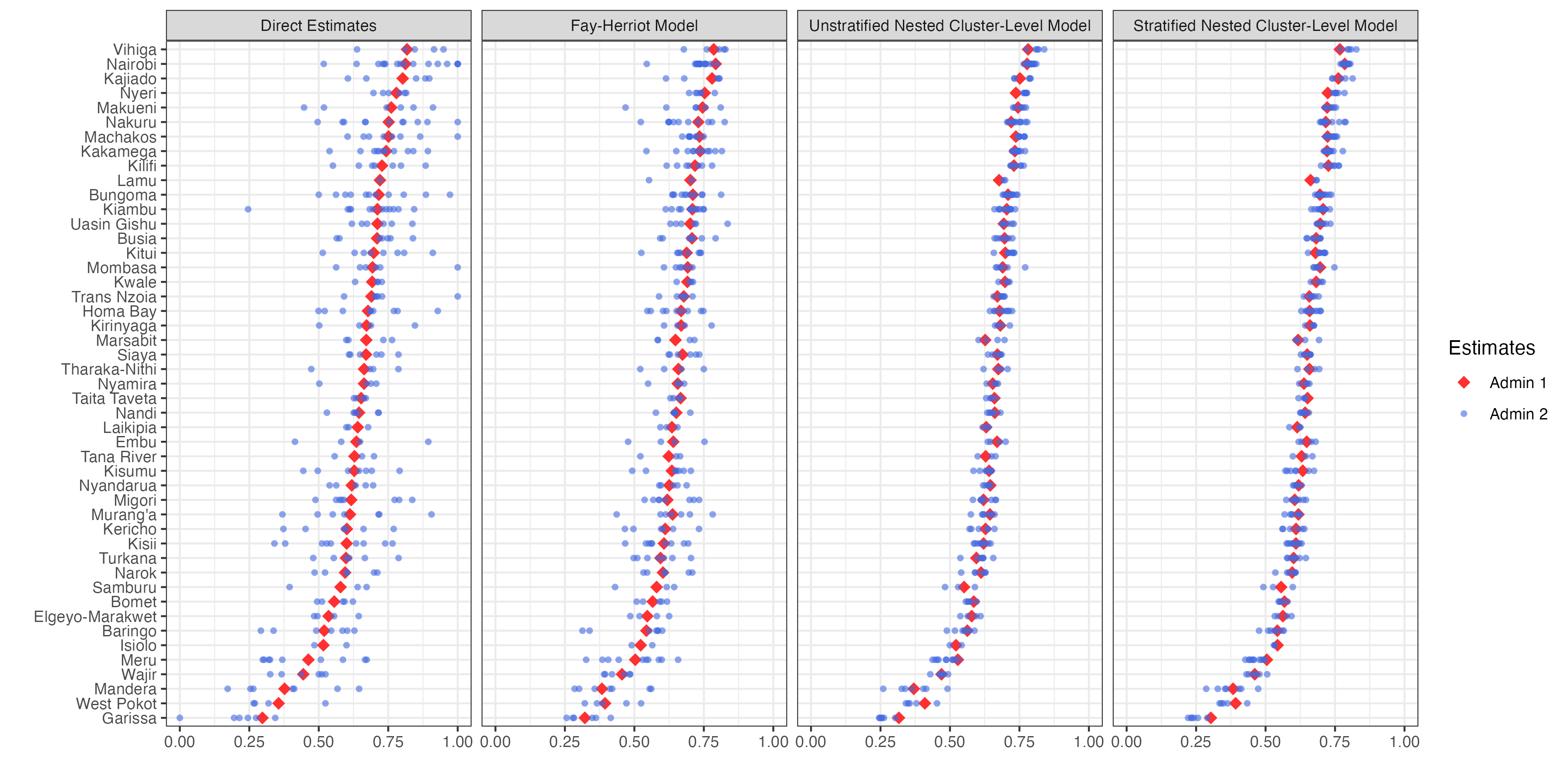}
    \caption{Point estimates of prevalence from different models. Blue points correspond to estimates from Admin-2 level models and red points correspond to estimates from Admin-1 level models. The Admin-2 level estimates are grouped by their Admin-1 level membership. For the smoothing models, the Admin-1 estimates are produced by fitting a separate model and are not directly aggregated from Admin-2 level results.}
    \label{fig:overlay}
\end{figure}

\rev{In terms of model comparison, scatter plots such as Figure \ref{fig:cov-scatter} and \ref{fig:shrinkage} are useful in identifying potential over-smoothing. Figure \ref{fig:overlay} further provides a more concise summary of estimates at different levels by combining point estimates at both Admin-1 and Admin-2 levels.} The plot provides a useful visual check of whether Admin-2 level models lead to estimates that are reasonably consistent with the estimates based on Admin-1 level models. The shrinkage of Admin-2 estimates, however, should not be interpreted as a lack of variation within any given Admin-1 areas, but rather it indicates the lack of information in the data that allows us to capture such variation reliably. Similar cautions should be taken in interpreting the `flat' map of point estimates in Figure \ref{fig:map-comparison2}.

\begin{figure}[!ht]
    \centering
    \includegraphics[width=\linewidth]{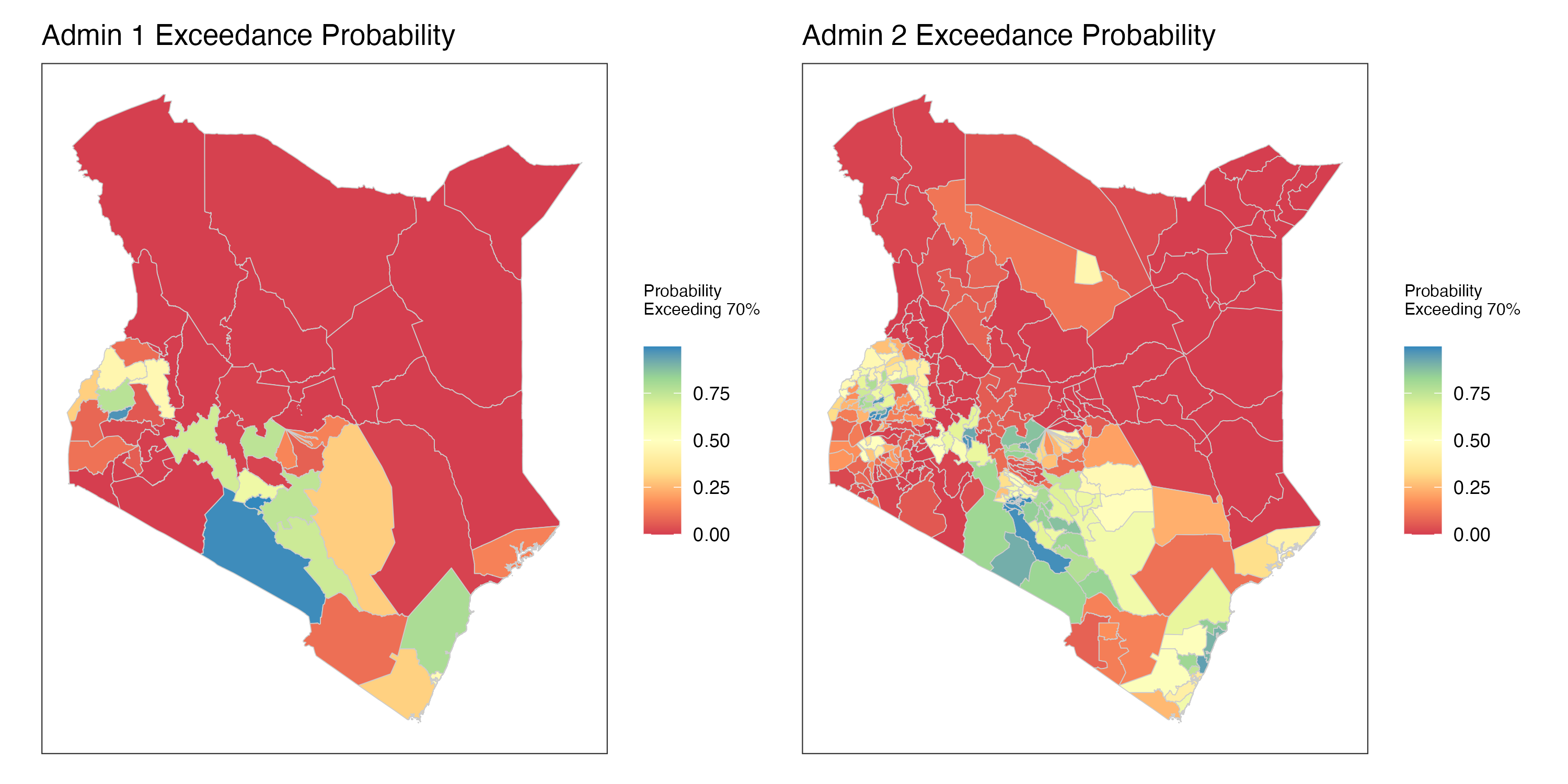}
    \caption{Probability of prevalence exceeding 70\% in each Admin-1 area (Left) and Admin-2 area (Right) using the nested stratified cluster-level model.}
    \label{fig:exceed}
\end{figure}

Lastly, the posterior distribution of the prevalence  are usually the input to compute other quantities of interest. For example, exceedance probabilities, that is, the probability that the prevalence of a given area exceeds certain thresholds, are often used to assess progress towards specific targets. Figure \ref{fig:exceed} shows the posterior probability of ANC4+ coverage exceeding the 70\% goal \citep{macharia2022spatial}, for each Admin-1 and Admin-2 area under the stratified nested cluster-level model. The striking band of red in the north and east is largely due to the great rurality of these regions of Kenya (see Figure \ref{fig:cluster-map}), and the general lack of health facilities (where women would travel for ANC visits) and increased travel times, which are an obvious deterrent. 

\section{Discussion} \label{sec:discuss}
In this paper, we propose a general workflow for mapping the prevalence of a binary outcome using household survey data in LMICs. The workflow is illustrated through an example estimating the subnational ANC4+ coverage in Kenya. 
The workflow includes steps to assess data availability and how they are collected, fitting a series of prevalence mapping models, model evaluation and comparison, and visualization and summarization. 
Throughout the discussion, we focus on the comparison of strengths and limitations across models, as well as potential adaptations to other contexts with varying data availability.

There are tasks in each of the steps where clear-cut recommendations cannot be provided.  Careful analysis and reasoning remain indispensable, and one must never forget the context of the data in question, with local knowledge being invaluable. \rev{In addition, the survey considered in this paper includes a reasonable number of samples. For smaller surveys, mapping the prevalence at fine geographic resolution can be more challenging and require careful model development, checking, and evaluation}. We emphasize that this workflow should be treated as a baseline analysis pipeline to assist analysts in further adaptations and extensions.

There are many open questions and challenges in prevalence mapping.
On the methodological side, more work needs to be done to connect the SAE  and geostatistical approaches. The stratified cluster-level model in our workflow is one example of utilizing flexible spatial models at the unit level while accounting for the sampling design. Model evaluation and comparison frameworks are ever more critical with these flexible models. The workflow discussed in this paper focuses on one binary indicator from a single household survey. The same principles extend to modeling more complex indicators, joint modeling of multiple related indicators, and modeling multiple surveys, e.g., mapping child mortality in space and time \citep{mercer:etal:15}. More complex models are needed to borrow information across the additional dimensions effectively. 
Finally, a major goal of this paper on workflow is to enable the prevalence mapping models and tools to be accessible to researchers and analysts in LMICs so that they can analyze their own data to answer their specific questions of priority, instead of relying on practically irreproducible metrics produced by prominent institutions. Understanding the limitations of the models and the uncertainties of the estimates is critical in designing better and more targeted data collection schemes in LMICs.

\bibliographystyle{chicago}
\bibliography{spatial}

\begin{thebibliography}{}

\bibitem[\protect\citeauthoryear{Altay, Paige, Riebler, and Fuglstad}{Altay
  et~al.}{2025}]{altay2025impact}
Altay, U., J.~Paige, A.~Riebler, and G.-A. Fuglstad (2025).
\newblock Impact of jittering on raster-and distance-based geostatistical
  analyses of {DHS} data.
\newblock {\em Statistical Modelling\/}~{\em 25\/}(1), 55--74.

\bibitem[\protect\citeauthoryear{Besag, York, and Molli\'{e}}{Besag
  et~al.}{1991}]{besag:york:mollie:91}
Besag, J., J.~York, and A.~Molli\'{e} (1991).
\newblock Bayesian image restoration with two applications in spatial
  statistics.
\newblock {\em Annals of the Institute of Statistics and Mathematics\/}~{\em
  43}, 1--59.

\bibitem[\protect\citeauthoryear{Blangiardo and Cameletti}{Blangiardo and
  Cameletti}{2015}]{blangiardo:cameletti:15}
Blangiardo, M. and M.~Cameletti (2015).
\newblock {\em Spatial and Spatio-Temporal Bayesian Models with R-INLA}.
\newblock John Wiley and Sons.

\bibitem[\protect\citeauthoryear{Burris and Hoff}{Burris and
  Hoff}{2020}]{burris2020exact}
Burris, K.~C. and P.~D. Hoff (2020).
\newblock Exact adaptive confidence intervals for small areas.
\newblock {\em Journal of Survey Statistics and Methodology\/}~{\em 8},
  206--230.

\bibitem[\protect\citeauthoryear{Burstein, Wang, Reiner~Jr, and Hay}{Burstein
  et~al.}{2018}]{burstein:etal:18}
Burstein, R., H.~Wang, R.~C. Reiner~Jr, and S.~I. Hay (2018).
\newblock Development and validation of a new method for indirect estimation of
  neonatal, infant, and child mortality trends using summary birth histories.
\newblock {\em PLoS Medicine\/}~{\em 15}, e1002687.

\bibitem[\protect\citeauthoryear{Chandra, Chambers, and Salvati}{Chandra
  et~al.}{2019}]{chandra2019small}
Chandra, H., R.~Chambers, and N.~Salvati (2019).
\newblock Small area estimation of survey weighted counts under aggregated
  level spatial model.
\newblock {\em Survey Methodology\/}~{\em 45\/}(1), 31--59.

\bibitem[\protect\citeauthoryear{Chi, Fang, Chatterjee, and Blumenstock}{Chi
  et~al.}{2022}]{chi2022microestimates}
Chi, G., H.~Fang, S.~Chatterjee, and J.~E. Blumenstock (2022).
\newblock Microestimates of wealth for all low-and middle-income countries.
\newblock {\em Proceedings of the National Academy of Sciences\/}~{\em
  119\/}(3), e2113658119.

\bibitem[\protect\citeauthoryear{Chung and Datta}{Chung and
  Datta}{2020}]{chung:datta:20}
Chung, H.~C. and G.~S. Datta (2020).
\newblock Bayesian hierarchical spatial models for small area estimation.
\newblock Technical report, Center for Statistical Research \& Methodology,
  U.S. Census Bureau.

\bibitem[\protect\citeauthoryear{Cloutier and Langlet}{Cloutier and
  Langlet}{2014}]{cloutier2014aboriginal}
Cloutier, E.~C. and {\'E}.~Langlet (2014).
\newblock {\em Aboriginal {P}eoples {S}urvey, 2012: {C}oncepts and {M}ethods
  {G}uide}.
\newblock Statistics Canada=Statistique Canada.

\bibitem[\protect\citeauthoryear{Croft, Allen, and Zachary}{Croft
  et~al.}{2023}]{dhsguide}
Croft, T.~N., C.~K. Allen, and B.~W. Zachary (2023).
\newblock Guide to {DHS} statistics.
\newblock {\em Rockville, Maryland, USA: ICF\/}.

\bibitem[\protect\citeauthoryear{Didan, Munoz, Solano, Huete, et~al.}{Didan
  et~al.}{2015}]{didan2015modis}
Didan, K., A.~B. Munoz, R.~Solano, A.~Huete, et~al. (2015).
\newblock {MODIS} vegetation index user’s guide ({MOD13} series).
\newblock {\em University of Arizona: Vegetation Index and Phenology
  Lab\/}~{\em 35}, 2--33.

\bibitem[\protect\citeauthoryear{Diggle and Giorgi}{Diggle and
  Giorgi}{2016}]{diggle:giorgi:16}
Diggle, P. and E.~Giorgi (2016).
\newblock Model-based geostatistics for prevalence mapping in low-resource
  settings.
\newblock {\em Journal of the American Statistical Association\/}~{\em 111},
  1096--1120.

\bibitem[\protect\citeauthoryear{Diggle and Giorgi}{Diggle and
  Giorgi}{2019}]{diggle:giorgi:19}
Diggle, P.~J. and E.~Giorgi (2019).
\newblock {\em Model-based Geostatistics for Global Public Health: Methods and
  Applications}.
\newblock Chapman and Hall/CRC.

\bibitem[\protect\citeauthoryear{Dong, Li, Wu, Boskovic, and Wakefield}{Dong
  et~al.}{2024}]{surveyPrev}
Dong, Q., Z.~R. Li, Y.~Wu, A.~Boskovic, and J.~Wakefield (2024).
\newblock {\em {surveyPrev}: Mapping the Prevalence of Binary Indicators using
  Survey Data in Small Areas}.
\newblock R package version 1.1.0.

\bibitem[\protect\citeauthoryear{Dong and Wakefield}{Dong and
  Wakefield}{2021}]{dong:wakefield:21}
Dong, T. and J.~Wakefield (2021).
\newblock Modeling and presentation of health and demographic indicators in a
  low- and middle-income countries context.
\newblock {\em Vaccine\/}~{\em 39}, 2584--2594.

\bibitem[\protect\citeauthoryear{Fay and Herriot}{Fay and
  Herriot}{1979}]{fay:herriot:79}
Fay, R. and R.~Herriot (1979).
\newblock Estimates of income for small places: an application of
  {James--Stein} procedure to census data.
\newblock {\em Journal of the American Statistical Association\/}~{\em 74},
  269--277.

\bibitem[\protect\citeauthoryear{Franco and Bell}{Franco and
  Bell}{2013}]{franco_applying_2013}
Franco, C. and W.~R. Bell (2013).
\newblock Applying {bivariate} {binomial}/{logit} {normal} {models} to {small}
  {area} {estimation}.
\newblock In {\em Proceedings of the {American} {Statistical} {Association},
  {Survey} {Research} {Section}}, pp.\  690--702.

\bibitem[\protect\citeauthoryear{Fuglstad, Hem, Knight, Rue, and
  Riebler}{Fuglstad et~al.}{2019}]{fuglstad:etal:19b}
Fuglstad, G.-A., I.~G. Hem, A.~Knight, H.~Rue, and A.~Riebler (2019).
\newblock Intuitive principle-based priors for attributing variance in additive
  model structures.
\newblock {\em arXiv preprint arXiv:1902.00242\/}.

\bibitem[\protect\citeauthoryear{Gadiaga, Abbott, Chamberlain, Lazar, Darin,
  and Tatem}{Gadiaga et~al.}{2023}]{kenyapop22}
Gadiaga, A.~N., T.~J. Abbott, H.~Chamberlain, A.~N. Lazar, E.~Darin, and A.~J.
  Tatem (2023).
\newblock Census disaggregated gridded population estimates for {K}enya (2022),
  version 2.0.

\bibitem[\protect\citeauthoryear{Gao and Wakefield}{Gao and
  Wakefield}{2023}]{gao2023spatial}
Gao, P.~A. and J.~Wakefield (2023).
\newblock A spatial variance-smoothing area level model for small area
  estimation of demographic rates.
\newblock {\em International Statistical Review\/}~{\em 91}, 493--510.

\bibitem[\protect\citeauthoryear{Gaughan, Oda, Sorichetta, Stevens, Bondarenko,
  Bun, Krauser, Yetman, and Nghiem}{Gaughan
  et~al.}{2019}]{gaughan2019evaluating}
Gaughan, A.~E., T.~Oda, A.~Sorichetta, F.~R. Stevens, M.~Bondarenko, R.~Bun,
  L.~Krauser, G.~Yetman, and S.~V. Nghiem (2019).
\newblock Evaluating nighttime lights and population distribution as proxies
  for mapping anthropogenic {CO2} emission in {Vietnam, Cambodia and Laos}.
\newblock {\em Environmental Research Communications\/}~{\em 1\/}(9), 091006.

\bibitem[\protect\citeauthoryear{Giorgi, Fronterr{\`e}, Macharia, Alegana,
  Snow, and Diggle}{Giorgi et~al.}{2021}]{giorgi2021model}
Giorgi, E., C.~Fronterr{\`e}, P.~M. Macharia, V.~A. Alegana, R.~W. Snow, and
  P.~J. Diggle (2021).
\newblock Model building and assessment of the impact of covariates for disease
  prevalence mapping in low-resource settings: to explain and to predict.
\newblock {\em Journal of The Royal Society Interface\/}~{\em 18}, 20210104.

\bibitem[\protect\citeauthoryear{{Global Administrative Areas}}{{Global
  Administrative Areas}}{2022}]{gadm}
{Global Administrative Areas} (2022).
\newblock {GADM} database of global administrative areas, version 4.1.

\bibitem[\protect\citeauthoryear{H{\'a}jek}{H{\'a}jek}{1971}]{hajek:71}
H{\'a}jek, J. (1971).
\newblock Discussion of, ``{A}n essay on the logical foundations of survey
  sampling, part {I}'', by {D}. {B}asu.
\newblock In V.~Godambe and D.~Sprott (Eds.), {\em Foundations of Statistical
  Inference}. Toronto: Holt, Rinehart and Winston.

\bibitem[\protect\citeauthoryear{Hoeting, Davis, Merton, and Thompson}{Hoeting
  et~al.}{2006}]{hoeting2006model}
Hoeting, J.~A., R.~A. Davis, A.~A. Merton, and S.~E. Thompson (2006).
\newblock Model selection for geostatistical models.
\newblock {\em Ecological Applications\/}~{\em 16\/}(1), 87--98.

\bibitem[\protect\citeauthoryear{Khan and Hancioglu}{Khan and
  Hancioglu}{2019}]{khan2019multiple}
Khan, S. and A.~Hancioglu (2019).
\newblock Multiple indicator cluster surveys: delivering robust data on
  children and women across the globe.
\newblock {\em Studies in Family Planning\/}~{\em 50}, 279--286.

\bibitem[\protect\citeauthoryear{KNBS and ICF}{KNBS and
  ICF}{2023}]{knbs2023kenya}
KNBS and ICF (2023).
\newblock {Kenya Demographic and Health Survey 2022}: volume 1.

\bibitem[\protect\citeauthoryear{Krainski, G{\'o}mez-Rubio, Bakka, Lenzi,
  Castro-Camilo, Simpson, Lindgren, and Rue}{Krainski
  et~al.}{2018}]{krainski:etal:18}
Krainski, E.~T., V.~G{\'o}mez-Rubio, H.~Bakka, A.~Lenzi, D.~Castro-Camilo,
  D.~Simpson, F.~Lindgren, and H.~Rue (2018).
\newblock {\em Advanced Spatial Modeling with Stochastic Partial Differential
  Equations Using R and INLA}.
\newblock Chapman and Hall/CRC.

\bibitem[\protect\citeauthoryear{Li, Hsiao, Godwin, Martin, Wakefield, and
  Clark}{Li et~al.}{2019}]{li:etal:19}
Li, Z.~R., Y.~Hsiao, J.~Godwin, B.~D. Martin, J.~Wakefield, and S.~J. Clark
  (2019).
\newblock Changes in the spatial distribution of the under five mortality rate:
  small-area analysis of 122 {DHS} surveys in 262 subregions of 35 countries in
  {A}frica.
\newblock {\em PLoS One\/}~{\em 14}, e0210645.

\bibitem[\protect\citeauthoryear{Liu, Lahiri, and Kalton}{Liu
  et~al.}{2014}]{liu_hierarchical_2014}
Liu, B., P.~Lahiri, and G.~Kalton (2014).
\newblock Hierarchical {Bayes} {Modeling} of {Survey}-{Weighted} {Small} {Area}
  {Proportions}.
\newblock {\em Survey Methodology\/}~{\em 40}, 1--13.

\bibitem[\protect\citeauthoryear{Lumley and Scott}{Lumley and
  Scott}{2014}]{lumley2014tests}
Lumley, T. and A.~Scott (2014).
\newblock Tests for regression models fitted to survey data.
\newblock {\em Australian \& New Zealand Journal of Statistics\/}~{\em
  56\/}(1), 1--14.

\bibitem[\protect\citeauthoryear{Macharia, Joseph, Nalwadda, Mwilike,
  Banke-Thomas, Benova, and Johnson}{Macharia
  et~al.}{2022}]{macharia2022spatial}
Macharia, P.~M., N.~K. Joseph, G.~K. Nalwadda, B.~Mwilike, A.~Banke-Thomas,
  L.~Benova, and O.~Johnson (2022).
\newblock Spatial variation and inequities in antenatal care coverage in
  {Kenya}, {Uganda} and mainland {Tanzania} using model-based geostatistics: a
  socioeconomic and geographical accessibility lens.
\newblock {\em BMC pregnancy and childbirth\/}~{\em 22}, 908.

\bibitem[\protect\citeauthoryear{Mayala, Dontamsetti, Fish, and Croft}{Mayala
  et~al.}{2019}]{mayala:etal:19}
Mayala, B., T.~Dontamsetti, T.~D. Fish, and T.~N. Croft (2019).
\newblock {\em Interpolation of {DHS} Survey Data at Subnational Administrative
  Level 2}.
\newblock DHS Spatial Analysis Reports No.~17. Rockville, Maryland, USA.

\bibitem[\protect\citeauthoryear{Mercer, Wakefield, Pantazis, Lutambi, Mosanja,
  and Clark}{Mercer et~al.}{2015}]{mercer:etal:15}
Mercer, L., J.~Wakefield, A.~Pantazis, A.~Lutambi, H.~Mosanja, and S.~Clark
  (2015).
\newblock Small area estimation of childhood mortality in the absence of vital
  registration.
\newblock {\em Annals of Applied Statistics\/}~{\em 9}, 1889--1905.

\bibitem[\protect\citeauthoryear{Merfeld, Chen, Lahiri, and Newhouse}{Merfeld
  et~al.}{2024}]{merfeld2024small}
Merfeld, J., H.~Chen, P.~Lahiri, and D.~Newhouse (2024).
\newblock Small area estimation with geospatial data: A primer.

\bibitem[\protect\citeauthoryear{Merfeld, Newhouse, Weber, and Lahiri}{Merfeld
  et~al.}{2022}]{merfeld2022combining}
Merfeld, J.~D., D.~L. Newhouse, M.~Weber, and P.~Lahiri (2022).
\newblock Combining survey and geospatial data can significantly improve
  gender-disaggregated estimates of labor market outcomes.

\bibitem[\protect\citeauthoryear{Michal, Wakefield, Schmidt, Cavanaugh,
  Robinson, and Baumgartner}{Michal et~al.}{2023}]{michal2023small}
Michal, V., J.~Wakefield, A.~M. Schmidt, A.~Cavanaugh, B.~Robinson, and
  J.~Baumgartner (2023).
\newblock Small area estimation with random forests and the {LASSO}.
\newblock {\em arXiv preprint arXiv:2308.15180\/}.

\bibitem[\protect\citeauthoryear{Mohadjer, Rao, Liu, Krenzke, and
  de~Kerckhove}{Mohadjer et~al.}{2012}]{mohadjer_hierarchical_2012}
Mohadjer, L., J.~N.~K. Rao, B.~Liu, T.~Krenzke, and W.~V. de~Kerckhove (2012).
\newblock Hierarchical {Bayes} {Small} {Area} {Estimates} of {Adult} {Literacy}
  using {Unmatched} {Sampling} and {Linking} {Models}.
\newblock {\em Journal of the Indian Society of Agricultural Statistics\/},
  55--63.

\bibitem[\protect\citeauthoryear{Molina, Rao, and Guadarrama}{Molina
  et~al.}{2019}]{molina2019small}
Molina, I., J.~Rao, and M.~Guadarrama (2019).
\newblock Small area estimation methods for poverty mapping: a selective
  review.
\newblock {\em Statistics and Applications\/}~{\em 17}, 11--22.

\bibitem[\protect\citeauthoryear{Molina and Rao}{Molina and
  Rao}{2010}]{molina2010small}
Molina, I. and J.~N. Rao (2010).
\newblock Small area estimation of poverty indicators.
\newblock {\em Canadian Journal of Statistics\/}~{\em 38\/}(3), 369--385.

\bibitem[\protect\citeauthoryear{Osgood-Zimmerman, Millear, Stubbs, Shields,
  Pickering, Earl, Graetz, Kinyoki, Ray, Bhatt, Browne, Burstein, Cameron,
  Casey, Deshpande, Fullman, Gething, Gibson, Henry, Herrero, Krause,
  Letourneau, Levine, Liu, Longbottom, Mayala, Mosser, Noor, Pigott, Piwoz,
  Rao, Rawat, Reiner, Smith, Weiss, Wiens, Mokdad, Lim, Murray, Kassebaum, and
  Hay}{Osgood-Zimmerman et~al.}{2018}]{osgood:etal:18}
Osgood-Zimmerman, A., A.~I. Millear, R.~W. Stubbs, C.~Shields, B.~V. Pickering,
  L.~Earl, N.~Graetz, D.~K. Kinyoki, S.~E. Ray, S.~Bhatt, A.~Browne,
  R.~Burstein, E.~Cameron, D.~Casey, A.~Deshpande, N.~Fullman, P.~Gething,
  H.~Gibson, N.~Henry, M.~Herrero, L.~Krause, I.~Letourneau, A.~Levine, P.~Liu,
  J.~Longbottom, B.~Mayala, J.~Mosser, A.~Noor, D.~Pigott, E.~Piwoz, P.~Rao,
  R.~Rawat, R.~Reiner, D.~Smith, D.~Weiss, K.~Wiens, A.~Mokdad, S.~Lim,
  C.~Murray, N.~Kassebaum, and S.~Hay (2018).
\newblock Mapping child growth failure in {A}frica between 2000 and 2015.
\newblock {\em Nature\/}~{\em 555}, 41.

\bibitem[\protect\citeauthoryear{Osgood-Zimmerman and
  Wakefield}{Osgood-Zimmerman and
  Wakefield}{2023}]{osgood-zimmerman:wakefield:19}
Osgood-Zimmerman, A. and J.~Wakefield (2023).
\newblock A statistical review of template model builder: A flexible tool for
  spatial modelling.
\newblock {\em International Statistical Review\/}~{\em 91}, 318--342.

\bibitem[\protect\citeauthoryear{Otto and Bell}{Otto and
  Bell}{1995}]{otto:bell:95}
Otto, M.~C. and W.~R. Bell (1995).
\newblock Sampling error modelling of poverty and income statistics for states.
\newblock In {\em American Statistical Association, Proceedings of the Section
  on Government Statistics}, pp.\  160--165.

\bibitem[\protect\citeauthoryear{Pfeffermann}{Pfeffermann}{2013}]{pfefferman:13}
Pfeffermann, D. (2013).
\newblock New important developments in small area estimation.
\newblock {\em Statistical Science\/}~{\em 28}, 40--68.

\bibitem[\protect\citeauthoryear{{R Core Team}}{{R Core Team}}{2024}]{Rlang}
{R Core Team} (2024).
\newblock {\em R: A Language and Environment for Statistical Computing}.
\newblock Vienna, Austria: R Foundation for Statistical Computing.

\bibitem[\protect\citeauthoryear{Rao and Molina}{Rao and
  Molina}{2015}]{rao:molina:15}
Rao, J. and I.~Molina (2015).
\newblock {\em Small Area Estimation, Second Edition}.
\newblock New York: John Wiley.

\bibitem[\protect\citeauthoryear{Riebler, S{\o}rbye, Simpson, and Rue}{Riebler
  et~al.}{2016}]{riebler:etal:16}
Riebler, A., S.~S{\o}rbye, D.~Simpson, and H.~Rue (2016).
\newblock An intuitive {B}ayesian spatial model for disease mapping that
  accounts for scaling.
\newblock {\em Statistical Methods in Medical Research\/}~{\em 25}, 1145--1165.

\bibitem[\protect\citeauthoryear{{Roll Back Malaria}}{{Roll Back
  Malaria}}{2005}]{malaria2005malaria}
{Roll Back Malaria} (2005).
\newblock {Malaria Indicator Survey}: Basic documentation for survey design and
  implementation.
\newblock {\em {Calverton MD: RBM Monitoring and Evaluation Research Group,
  WHO, UNICEF, Measure Evaluation, US Centers for Disease Control and
  Prevention}\/}.

\bibitem[\protect\citeauthoryear{Rom{\'a}n, Wang, Sun, Kalb, Miller, Molthan,
  Schultz, Bell, Stokes, Pandey, et~al.}{Rom{\'a}n
  et~al.}{2018}]{roman2018nasa}
Rom{\'a}n, M.~O., Z.~Wang, Q.~Sun, V.~Kalb, S.~D. Miller, A.~Molthan,
  L.~Schultz, J.~Bell, E.~C. Stokes, B.~Pandey, et~al. (2018).
\newblock Nasa's black marble nighttime lights product suite.
\newblock {\em Remote Sensing of Environment\/}~{\em 210}, 113--143.

\bibitem[\protect\citeauthoryear{Rue, Martino, and Chopin}{Rue
  et~al.}{2009}]{rue:etal:09}
Rue, H., S.~Martino, and N.~Chopin (2009).
\newblock Approximate {B}ayesian inference for latent {G}aussian models using
  integrated nested {L}aplace approximations (with discussion).
\newblock {\em Journal of the Royal Statistical Society, Series B\/}~{\em 71},
  319--392.

\bibitem[\protect\citeauthoryear{Saha, Das, Baffour, and Chandra}{Saha
  et~al.}{2023}]{saha2023small}
Saha, U.~R., S.~Das, B.~Baffour, and H.~Chandra (2023).
\newblock Small area estimation of age-specific and total fertility rates in
  {B}angladesh.
\newblock {\em Spatial Demography\/}~{\em 11\/}(1), 2.

\bibitem[\protect\citeauthoryear{Simpson, Rue, Riebler, Martins, and
  S{\o}rbye}{Simpson et~al.}{2017}]{simpson:etal:17}
Simpson, D., H.~Rue, A.~Riebler, T.~Martins, and S.~S{\o}rbye (2017).
\newblock Penalising model component complexity: A principled, practical
  approach to constructing priors (with discussion).
\newblock {\em Statistical Science\/}~{\em 32}, 1--28.

\bibitem[\protect\citeauthoryear{S{\o}rbye and Rue}{S{\o}rbye and
  Rue}{2017}]{sorbye2017penalised}
S{\o}rbye, S.~H. and H.~Rue (2017).
\newblock Penalised complexity priors for stationary autoregressive processes.
\newblock {\em Journal of Time Series Analysis\/}~{\em 38\/}(6), 923--935.

\bibitem[\protect\citeauthoryear{Tatem}{Tatem}{2017}]{tatem2017worldpop}
Tatem, A.~J. (2017).
\newblock World{P}op, open data for spatial demography.
\newblock {\em Scientific data\/}~{\em 4}.

\bibitem[\protect\citeauthoryear{Torabi and Rao}{Torabi and
  Rao}{2008}]{torabi2008small}
Torabi, M. and J.~Rao (2008).
\newblock Small area estimation under a two-level model.
\newblock {\em Survey Methodology\/}~{\em 34\/}(1), 11.

\bibitem[\protect\citeauthoryear{Tzavidis, Zhang, Luna, Schmid, and
  Rojas-Perilla}{Tzavidis et~al.}{2018}]{tzavidis:etal:18}
Tzavidis, N., L.-C. Zhang, A.~Luna, T.~Schmid, and N.~Rojas-Perilla (2018).
\newblock From start to finish: a framework for the production of small area
  official statistics.
\newblock {\em Journal of the Royal Statistical Society: Series A\/}~{\em 181},
  927--979.

\bibitem[\protect\citeauthoryear{{United Nations}}{{United
  Nations}}{2015}]{sdgsWeb}
{United Nations} (2015).
\newblock {\em Sustainable Development Goals}.
\newblock \url{https://sdgs.un.org/2030agenda}.

\bibitem[\protect\citeauthoryear{USAID}{USAID}{2019}]{dhs}
USAID (2019).
\newblock {\em Demographic and Health Surveys}.
\newblock \url{http://www.dhsprogram.com}: {United States Agency for
  International Development}.

\bibitem[\protect\citeauthoryear{Utazi, Thorley, Alegana, Ferrari, Takahashi,
  Metcalf, Lessler, and Tatem}{Utazi et~al.}{2018}]{utazi:etal:18}
Utazi, C.~E., J.~Thorley, V.~A. Alegana, M.~J. Ferrari, S.~Takahashi, C.~J.~E.
  Metcalf, J.~Lessler, and A.~J. Tatem (2018).
\newblock High resolution age-structured mapping of childhood vaccination
  coverage in low and middle income countries.
\newblock {\em Vaccine\/}~{\em 36}, 1583--1591.

\bibitem[\protect\citeauthoryear{Wairoto, Joseph, Macharia, and Okiro}{Wairoto
  et~al.}{2020}]{wairoto2020determinants}
Wairoto, K.~G., N.~K. Joseph, P.~M. Macharia, and E.~A. Okiro (2020).
\newblock Determinants of subnational disparities in antenatal care
  utilisation: a spatial analysis of demographic and health survey data in
  {K}enya.
\newblock {\em BMC Health Services Research\/}~{\em 20}, 665.

\bibitem[\protect\citeauthoryear{Wakefield, Fuglstad, Riebler, Godwin, Wilson,
  and Clark}{Wakefield et~al.}{2019}]{wakefield:etal:19}
Wakefield, J., G.-A. Fuglstad, A.~Riebler, J.~Godwin, K.~Wilson, and S.~Clark
  (2019).
\newblock Estimating under five mortality in space and time in a developing
  world context.
\newblock {\em Statistical Methods in Medical Research\/}~{\em 28}, 2614--2634.

\bibitem[\protect\citeauthoryear{Wakefield, Gao, Fuglstad, and Li}{Wakefield
  et~al.}{2025}]{wakefield2020two}
Wakefield, J., P.~A. Gao, G.-A. Fuglstad, and Z.~R. Li (2025).
\newblock The two cultures for prevalence mapping: small area estimation and
  model-based geostatistics.
\newblock {\em Statistical Science\/}.

\bibitem[\protect\citeauthoryear{Wakefield, Jitong, and Wu}{Wakefield
  et~al.}{2025}]{wakefield:etal:25}
Wakefield, J., J.~Jitong, and Y.~Wu (2025).
\newblock Variance adjustment in the {F}ay-{H}erriot model using a pseudo
  prior.
\newblock {\em Manuscript under Preparation\/}.

\bibitem[\protect\citeauthoryear{Wakefield, Okonek, and Pedersen}{Wakefield
  et~al.}{2020}]{wakefield2020small}
Wakefield, J., T.~Okonek, and J.~Pedersen (2020).
\newblock Small area estimation for disease prevalence mapping.
\newblock {\em International Statistical Review\/}~{\em 88}, 398--418.

\bibitem[\protect\citeauthoryear{Weiss, Nelson, Vargas-Ruiz, Gligori{\'c},
  Bavadekar, Gabrilovich, Bertozzi-Villa, Rozier, Gibson, Shekel, et~al.}{Weiss
  et~al.}{2020}]{weiss2020global}
Weiss, D., A.~Nelson, C.~Vargas-Ruiz, K.~Gligori{\'c}, S.~Bavadekar,
  E.~Gabrilovich, A.~Bertozzi-Villa, J.~Rozier, H.~Gibson, T.~Shekel, et~al.
  (2020).
\newblock Global maps of travel time to healthcare facilities.
\newblock {\em Nature Medicine\/}~{\em 26}, 1835--1838.

\bibitem[\protect\citeauthoryear{{World Health Organization}}{{World Health
  Organization}}{2002}]{world2002antenatal}
{World Health Organization} (2002).
\newblock {WHO} antenatal care randomized trial: manual for the implementation
  of the new model.
\newblock Technical report, World Health Organization.

\bibitem[\protect\citeauthoryear{Wu, Li, Mayala, Wang, Gao, Paige, Fuglstad,
  Moe, Godwin, Donohue, Janocha, Croft, and Wakefield}{Wu
  et~al.}{2021}]{wu:etal:21}
Wu, Y., Z.~R. Li, B.~Mayala, H.~Wang, P.~Gao, J.~Paige, G.-A. Fuglstad, C.~Moe,
  J.~Godwin, R.~Donohue, B.~Janocha, T.~Croft, and J.~Wakefield (2021).
\newblock {\em Spatial Modeling for Subnational Administrative level 2
  Small-Area Estimation}.
\newblock DHS Spatial Analysis Reports No.~21. Rockville, Maryland, USA.

\bibitem[\protect\citeauthoryear{Wu and Wakefield}{Wu and
  Wakefield}{2024}]{wu2024modelling}
Wu, Y. and J.~Wakefield (2024).
\newblock Modelling urban/rural fractions in low-and middle-income countries.
\newblock {\em Journal of the Royal Statistical Society Series A: Statistics in
  Society\/}, qnae003.

\end{thebibliography}

\end{document}